\documentclass[
reprint,
superscriptaddress,
amsmath,
amssymb,
aps,
prb,
floatfix,
]{revtex4-2}
\usepackage{graphicx}
\usepackage{hyperref}
\usepackage{xcolor}
\usepackage{multirow}

\begin{document}

\title{Structure, short-range order, and phase stability of the Al$_x$CrFeCoNi high-entropy alloy: Insights from a perturbative, DFT-based analysis}

\author{Christopher D. Woodgate}
\email{christopher.woodgate@bristol.ac.uk}
\affiliation{Department of Physics, University of Warwick, Coventry, CV4 7AL, United Kingdom}
\affiliation{H. H. Wills Physics Laboratory, University of Bristol, Royal Fort, Bristol, BS8 1TL, United Kingdom}
\author{George A. Marchant}
\email{George.Marchant@warwick.ac.uk}
\affiliation{Department of Physics, University of Warwick, Coventry, CV4 7AL, United Kingdom}
\affiliation{Department of Chemistry, University of Warwick, Coventry, CV4 7AL, United Kingdom}
\author{Livia B. P\'{a}rtay}
\email{Livia.Bartok-Partay@warwick.ac.uk}
\affiliation{Department of Chemistry, University of Warwick, Coventry, CV4 7AL, United Kingdom}
\author{Julie B. Staunton}
\email{J.B.Staunton@warwick.ac.uk}
\affiliation{Department of Physics, University of Warwick, Coventry, CV4 7AL, United Kingdom}

\begin{abstract}
We study the phase behaviour of the Al$_x$CrFeCoNi high-entropy alloy. Our approach is based on a perturbative analysis of the internal energy of the paramagnetic solid solution as evaluated within the Korringa-Kohn-Rostoker formulation of density functional theory, using the coherent potential approximation to average over disorder. Via application of a Landau-type linear response theory, we infer preferential chemical orderings directly. In addition, we recover a pairwise form of the alloy internal energy suitable for study via atomistic simulations, which in this work are performed using the nested sampling algorithm, which is well-suited for studying complex potential energy surfaces. When the underlying lattice is fcc, at low concentrations of Al, depending on the value of $x$, we predict either an L1$_2$ or D0$_{22}$ ordering emerging below approximately 1000~K. On the other hand, when the underlying lattice is bcc, consistent with experimental observations, we predict B2 ordering temperatures higher than the melting temperature of the alloy, confirming that this ordered phase forms directly from the melt. For both fcc and bcc systems, chemical orderings are dominated by Al moving to one sublattice, Ni and Co the other, while Cr and Fe remain comparatively disordered. On the bcc lattice, our atomistic modelling suggests eventual decomposition into B2 NiAl and Cr-rich phases. These results shed light on the fundamental physical origins of atomic ordering tendencies in these intriguing materials.
\end{abstract}

\date{October 3, 2024}

\maketitle

\section{Introduction}
\label{sec:introduction}

Since first reported experimentally in 2004~\cite{cantor_microstructural_2004, yeh_nanostructured_2004}, high-entropy alloys (HEAs) and, more generally, high-entropy materials, have attracted huge interest in the field of materials science~\cite{gao_high-entropy_2016, miracle_critical_2017, george_high-entropy_2019}. Not only have they been shown to exhibit a range of superior mechanical properties for applications, such as exceptional fracture resistance~\cite{gludovatz_fracture-resistant_2014} and damage tolerance~\cite{gludovatz_exceptional_2016}, but they are also of fundamental physical interest due to observed phenomena such as superconductivity~\cite{kozelj_discovery_2014}, quantum critical behaviour~\cite{sales_quantum_2016} and extreme Fermi-surface smearing~\cite{robarts_extreme_2020}.

One widely experimentally studied HEA is the Al$_x$CrFeCoNi system~\cite{kao_microstructure_2009}, which frequently exhibits coexistence between A1 (disordered face centred cubic, fcc) and B2 (ordered structure with an underlying body centred cubic, bcc, lattice) phases at intermediate values of $x$~\cite{bloomfield_phase_2022}. It is understood that varying Al content and precise control of processing conditions allow for the microstructure of this alloy to be tuned, resulting in improved mechanical properties for elevated temperature applications, such as in the nuclear, turbine and aerospace industries~\cite{xia_irradiation_2015, li_high-entropy_2017, praveen_highentropy_2018}. However, there remains debate in the literature about the equilibrium phase diagram of this system; some studies report as-cast microstructures~\cite{li_effect_2010, wang_effects_2012} which frequently contain inhomogeneities~\cite{manzoni_phase_2013}, while those studies studying phase equilibria report concerns over the long timescales required to achieve equilibrium~\cite{zhang_understanding_2016,  cieslak_phase_2018}. There is therefore a degree of uncertainty around the phase diagram of this system, which we believe warrants further study.

From the point of view of materials modelling, this class of materials presents a number of challenges for conventional modelling techniques~\cite{widom_modeling_2018, ferrari_frontiers_2020, ferrari_simulating_2023}. First, the large number of constituent elements at, in principle, arbitrary concentrations means that comparatively large simulation cells must be used to model systems with the correct overall composition. Second, these entropy-stabilised materials possess a huge degree of chemical (and often magnetic) disorder, meaning that the phase space is `large' and lots of configurations must be sampled for there to be confidence that results are well-converged and representative of the thermodynamic phases. Finally, there is a vast number of HEA compositions and, with new compositions being continually reported, it is not desirable to use computationally intensive methodologies which produce results specific to only one potential composition. Despite these challenges, there is now a range of methodologies in the materials modeller's `toolbox' suitable for examining the physics of these complex materials. These include large-scale density functional theory (DFT) studies~\cite{tamm_atomic-scale_2015}; interatomic potentials~\cite{farkas_model_2020, kim_interaction_2023, hasan_short-range_2024}, including machine-learned interatomic potentials~\cite{kostiuchenko_impact_2019, rosenbrock_machine-learned_2021, ghosh_short-range_2022} and cluster expansions~\cite{fernandez-caballero_short-range_2017}; and approaches based on effective medium theories such as the coherent potential approximation (CPA)~\cite{ruban_atomic_2004, singh_atomic_2015, schonfeld_local_2019, singh_tuning_2019}.

\begin{figure*}[t]
    \centering
    \includegraphics[width=\textwidth]{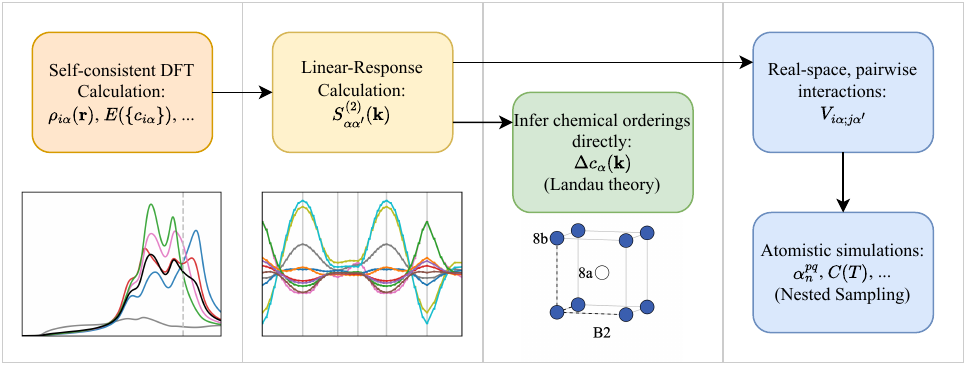}
    \caption{Visualisation of the workflow used in this work for studying the phase stability of the Al$_x$CrFeCoNi alloy, as discussed in Sections~\ref{sec:introduction} and \ref{sec:theory}.}
    \label{fig:workflow}
\end{figure*}

Our own methodology~\cite{khan_statistical_2016, woodgate_compositional_2022, woodgate_short-range_2023, woodgate_interplay_2023, woodgate_competition_2024, woodgate_modelling_2024, woodgate_integrated_nodate} for modelling the phase stability of these alloys falls into the last of the aforementioned categories, and is based on a perturbative analysis of the internal energy of the disordered alloy as evaluated within the CPA. Notably, compared to some alternative CPA-based schemes, we carefully and fully include the effects of charge rearrangement due to an applied chemical perturbation~\cite{khan_statistical_2016}, which turns out to be essential when modelling many HEA systems~\cite{woodgate_compositional_2022, woodgate_short-range_2023}. A schematic of the workflow is visualised in Fig.~\ref{fig:workflow}. In summary, the process begins by performing a self-consistent DFT calculation to evaluate the internal energy of the disordered solid solution. A perturbative analysis then examines the stability of this disordered phase to applied chemical fluctuations within a \textit{concentration wave} formalism, as pioneered for binary alloys by Khachaturyan~\cite{khachaturyan_ordering_1978} and Gyorffy and Stocks~\cite{gyorffy_concentration_1983}. Via application of a Landau-type theory, it is possible to infer chemical orderings directly, obtaining the preferred ordered structure, the temperature at which it emerges, and information about which chemical species preferentially sit on which sublattice. In addition to the results of the Landau theory, from this perturbative analysis it is also possible to recover a representation of the internal energy of the alloy in terms of effective pair interactions. This effective Hamiltonian can then be used in computationally inexpensive atomistic simulations to study the phase behaviour of a system below any initial ordering temperature in detail. Previously, we have made use of the Metropolis Monte Carlo algorithm for this purpose~\cite{woodgate_compositional_2022, woodgate_short-range_2023, woodgate_interplay_2023, woodgate_competition_2024, woodgate_modelling_2024, woodgate_integrated_nodate}, but in the present work we instead apply the nested sampling algorithm~\cite{NS_mat_review, NS_all_review}, a sampling technique which is fully predictive, truly unbiased, and capable of simultaneously exploring competing free energy basins. It is therefore well-suited to studying the complex phase behaviour of high-entropy alloys.

In this paper, to give fundamental physical insight into the experimentally observed phase behaviour of this system, we apply the above modelling approach to the Al$_x$CrFeCoNi system across a range of values of $x$. In alignment with experimental observations and our own calculations of the preferred lattice type, we simulate the system both when the underlying lattice is fcc (appropriate at low values of $x$) but also when the underlying lattice is bcc (appropriate at higher values of $x$). Our results enable a detailed interpretation of existing experimental data by predicting which chemical species partition onto which sublattice(s) when chemical orderings emerge in the system. Via application of the nested sampling algorithm, which uses a family of walkers to make a full exploration of the potential energy surface of a system across a range of temperatures, we are also able to examine the dominant atom-atom correlations in the system after any initial chemical ordering. Our results, therefore, facilitate detailed interpretation of experimental results, as well as providing insight into the physical origins of these chemical orderings.

\section{Results}
\label{sec:results}

\subsection{Underlying Crystal Lattice}

\begin{figure}[b]
    \centering
    \includegraphics[width=\linewidth]{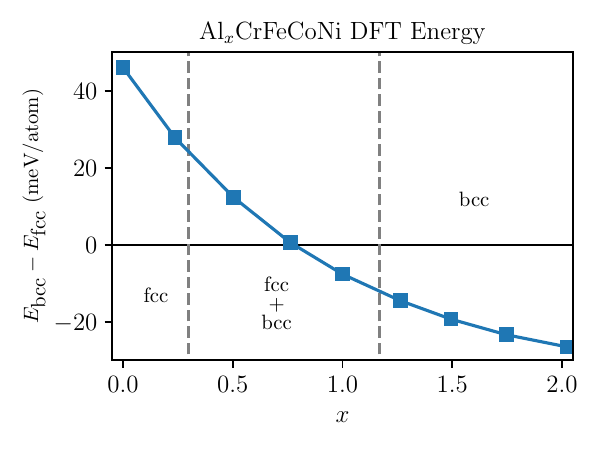}
    \caption{Difference in total DFT energy-per-atom between bcc and fcc structures for the disordered Al$_x$CrFeCoNi alloy as a function of $x$, as modelled within the KKR-CPA. Dashed lines indicate the ranges of $x$ where, from left to right experimentally the alloy is observed to form an fcc phase, a mixed fcc/bcc phase, and a pure bcc phase~\cite{kao_microstructure_2009, wang_effects_2012} It can be seen that the transition from the fcc to bcc structures observed experimentally is well-captured by the DFT data.}
    \label{fig:fcc_vs_bcc_energy}
\end{figure}

We begin by constructing the self-consistent, one-electron potentials of DFT~\cite{martin_electronic_2004}. Within the Korringa-Kohn-Rostoker (KKR) formulation of DFT~\cite{ebert_calculating_2011, faulkner_multiple_2018}, we use the coherent potential approximation (CPA)~\cite{soven_coherent-potential_1967, faulkner_calculating_1980, johnson_total-energy_1990} to model the electronic structure of the disordered solid solution. The CPA seeks to construct an effective medium of electronic scatterers whose overall scattering properties approximate those of the disordered solid solution. Rather than specifying an on-lattice configuration by a set of discrete site occupancies, $\{\xi_{i\alpha}\}$, where $\xi_{i\alpha}=1$ if site $i$ is occupied by an atom of species $\alpha$ and zero otherwise, the CPA works with \textit{partial} site occupancies,
\begin{equation}
    c_{i\alpha} = \langle \xi_{i\alpha} \rangle,
    \label{eq:partial_occupancies}
\end{equation}
where angle brackets denote an average over an appropriate ensemble of configurations. Above any chemical order-disorder transition temperature, these partial site occupancies, or \textit{site-wise concentrations}, are spatially homogeneous, \textit{i.e.}
\begin{equation}
    c_{i\alpha} \equiv c_\alpha,
    \label{eq:homogeneous_partial_occupancies}
\end{equation}
where $c_\alpha$ is the total (average) concentration of chemical species $\alpha$. In this homogeneous case, the CPA uses a self-consistency condition to construct a set of one-electron potentials whose average electronic scattering properties are reflective of those of the chemically disordered alloy~\cite{soven_coherent-potential_1967, faulkner_calculating_1980, ebert_calculating_2011, faulkner_multiple_2018}. Previous studies have verified that the CPA successfully captures many physical properties of HEAs, with examples including successful reproduction of smeared-out Fermi surfaces~\cite{robarts_extreme_2020}, bulk transport properties~\cite{jin_tailoring_2016}, and magnetic properties~\cite{billington_bulk_2020}. 

In addition to the consideration of compositional disorder, in alloys containing mid- to late- $3d$ transition metals such as Cr, Fe, Co, and Ni, it is important to treat the magnetic state appropriately~\cite{woodgate_interplay_2023}. The Al$_x$CrFeCoNi system is understood to have a Curie temperature of between approximately 200 and 400~K across the range of Al-concentrations considered in this work~\cite{kao_electrical_2011, huang_mechanism_2016}, which is lower than typical atomic ordering temperatures in these systems~\cite{bloomfield_phase_2022}. Therefore, throughout, we model these systems in their paramagnetic state within the disordered local moment (DLM) picture~\cite{pindor_disordered_1983, staunton_disordered_1984, gyorffy_first-principles_1985}, which represents a means by which to obtain the average electronic structure of a system in its paramagnetic state, sampling over all possible spin configurations, including those which are non-collinear. Across the range of $x$ considered in this paper we find that, when the underlying lattice is fcc, only Fe supports a local moment while, when the underlying lattice is bcc, we find that both Fe and Co support local moments.

Using the all-electron HUTSEPOT code~\cite{hoffmann_magnetic_2020}, we compute the total energy-per-atom of the disordered Al$_x$CrFeCoNi system in both disordered fcc (A1) and bcc (A2) structures as a function of $x$ in the range $0 \leq x \leq 2$, and calculate the difference in energy between the two structures. The results are shown in Fig.~\ref{fig:fcc_vs_bcc_energy}. Lattice parameters are set to be consistent with their experimental values~\cite{wang_effects_2012}. For values of $x$ where the lattice type is not observed experimentally ({\it e.g.} bcc, $x=0$; fcc, $x=2$) we perform a volume-conserving transformation from fcc to bcc structures (or vice-versa) to obtain a lattice parameter. Full details of these calculations are described in Sec.~\ref{sec:computational_details}. 

Dashed lines on the figure indicate the regions where, experimentally, the system is observed to form an fcc lattice, a mixed fcc/bcc phase, or purely the bcc lattice{, where the experimental data is taken from Ref.~\cite{kao_microstructure_2009}. These authors report the observed phases for the Al$_x$CrFeCoNi high-entropy alloy in two conditions: as-cast, and homogenised at 1100~$^\circ$C for 24~hrs. In the as-cast samples, it is found that fcc-bcc phase coexistence occurs in the region $0.45<x<0.88$, while in the homogenised samples the coexistence region is enlarged to $0.30<x<1.17$. We use the larger of these two windows to plot the dashed lines in Fig.~\ref{fig:fcc_vs_bcc_energy}.}

It can be seen that the KKR-CPA calculations successfully capture the transition of the underlying crystal lattice from fcc to bcc with increasing $x$. We deduce that the experimentally observed fcc$+$bcc coexistence region originates from the small difference in energy between the two structures in the region $0.5 \leq x \leq 1$, leading to competition between the two structures.

\subsection{Electronic Structure}

\begin{figure}[b]
    \centering
    \includegraphics[width=\linewidth]{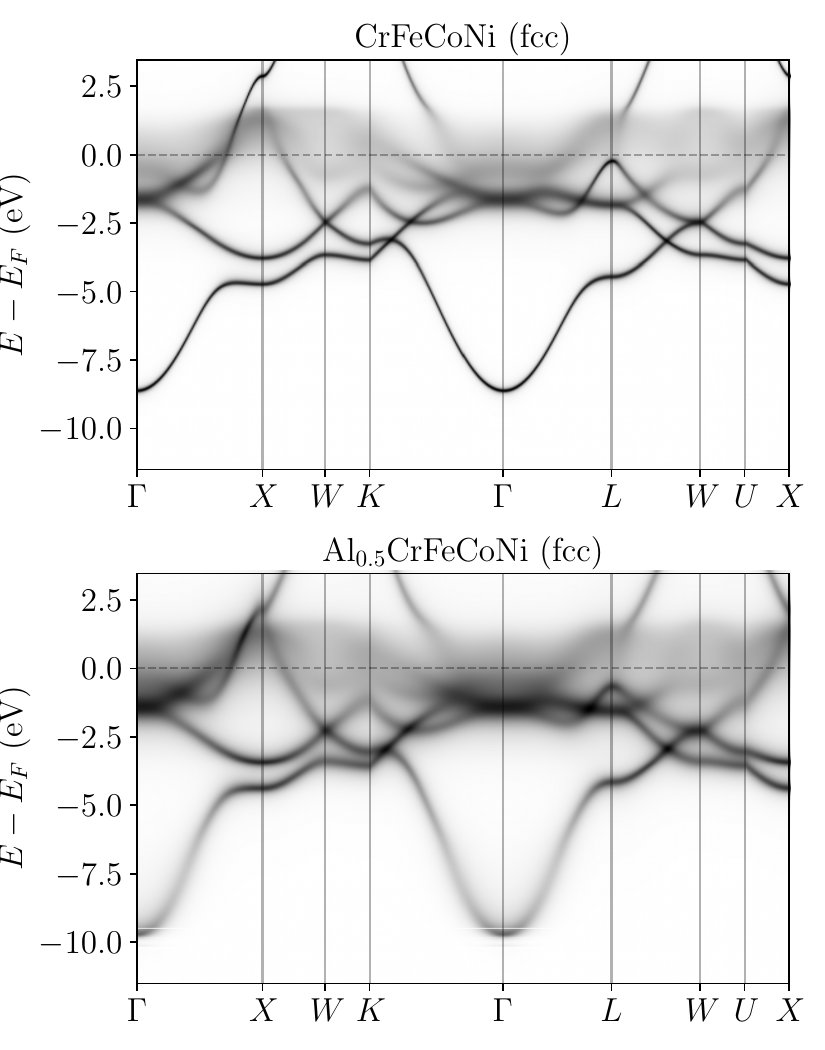}
    \caption{The Bloch spectral function (BSF) around the irreducible Brillouin zone (IBZ) for disordered (A1) Al$_x$CrFeCoNi on an underlying fcc lattice for $x=0$ and $x=0.5$. In both cases, the combination of chemical and magnetic disorder can be seen to smear out fine detail in the systems' narrow $3d$ bands. However, upon the addition of Al, the sharp, near-parabolic, $sp$-like bands are seen to be smeared and brought down to lower energies, indicative of the formation of hybridised bonding states.}
    \label{fig:bsf_comparison}
\end{figure}

\begin{figure*}[t]
    \centering
    \includegraphics[width=\textwidth]{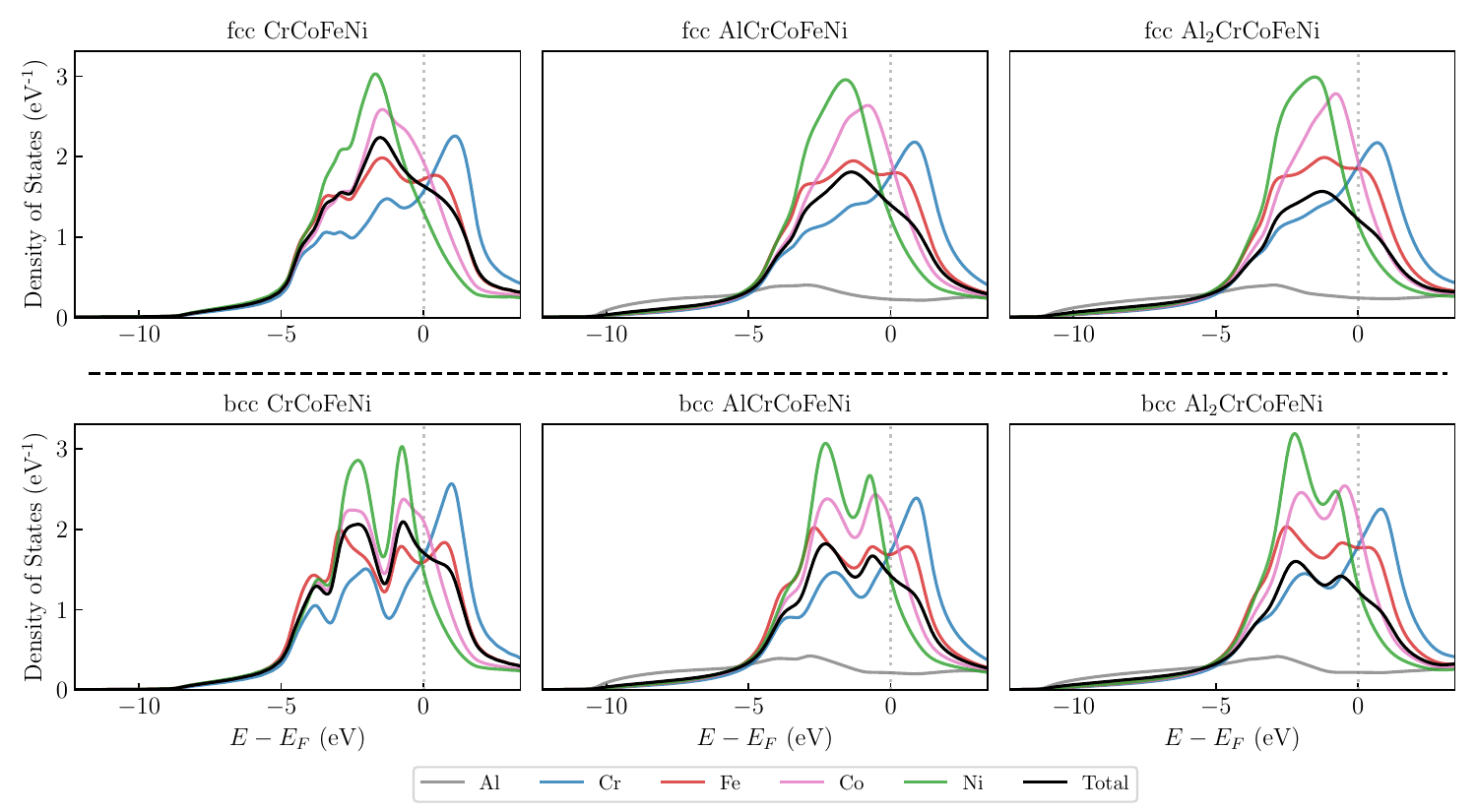}
    \caption{The total and species-resolved densities of states (DoS) for the disordered Al$_x$CrFeCoNi alloys for a selection of values of $x$ for both the fcc and bcc structures, as modelled within the KKR-CPA. The total DoS is given by the weighted average of the species-resolved curves, while the Fermi level is indicated by a grey, dashed line. The broad, $sp$-like band associated with Al can be seen running through and partially hybridising with the narrower $3d$ bands of the transition metals, as indicated by the local peaks in the species-resolved curve for Al.}
    \label{fig:dos_comparison}
\end{figure*}

Having performed calculations of the total energy-per-atom for each underlying lattice type, we examine the Bloch spectral function (BSF)~\cite{ebert_calculating_2011} and electronic density of states (DoS) around the Fermi level, $E_F$, to pull out salient features of the electronic `glue' of the system driving atomic ordering tendencies.  Fig.~\ref{fig:bsf_comparison} shows two indicative BSF plots for fcc Al$_x$CrFeCoNi, to demonstrate the impact Al has on the `bandstructure' of the system as modelled within the CPA.  Fig.~\ref{fig:dos_comparison} depicts the total and species- resolved DoS for the system for a range of $x$ values simulated in the A1 (disordered fcc, top row) and A2 (disordered bcc, bottom row) structures. Note that, although our calculations are spin-polarised, the DLM state averages over magnetic disorder for the paramagnetic state and it is therefore most meaningful to consider the total BSF and DoS.

Considering first the BSF plots of Fig.~\ref{fig:bsf_comparison}, we see that for both $x=0$ and $x=0.5$, the combination of compositional and magnetic disorder smears out the narrow, $3d$ bands around $E_F$. The degree of smearing of a particular band is associated with the electronic mean free path and, consequently, the lifetime of electronic states~\cite{ebert_calculating_2011, faulkner_multiple_2018}. Increased smearing means a shorter lifetime of electronic states, while a sharp, well-defined band means a long mean free path and consequently longer state lifetime. For $x=0$, our observed smearing of the $3d$ bands around $E_F$ is in alignment with experimental observations of the smeared-out Fermi surface of the CrFeCoNi system~\cite{robarts_extreme_2020}. However, we note that, without Al present in the system, (\textit{i.e.} the case $x=0$) the parabolic $sp$ bands at low energies remain relatively sharp, with minimal smearing. This picture is altered when Al is added to the system (bottom pane of Fig.~\ref{fig:bsf_comparison}), when it can be seen that these $sp$-like states are smeared and brought to lower energies, indicative of the formation of hybridised bonding states. In addition, the smearing of the $3d$ bands is increased, suggesting a degree of $p$-$d$ hybridisation between Al and the transition metals.

\begin{figure*}[t]
    \centering
    \includegraphics[width=\textwidth]{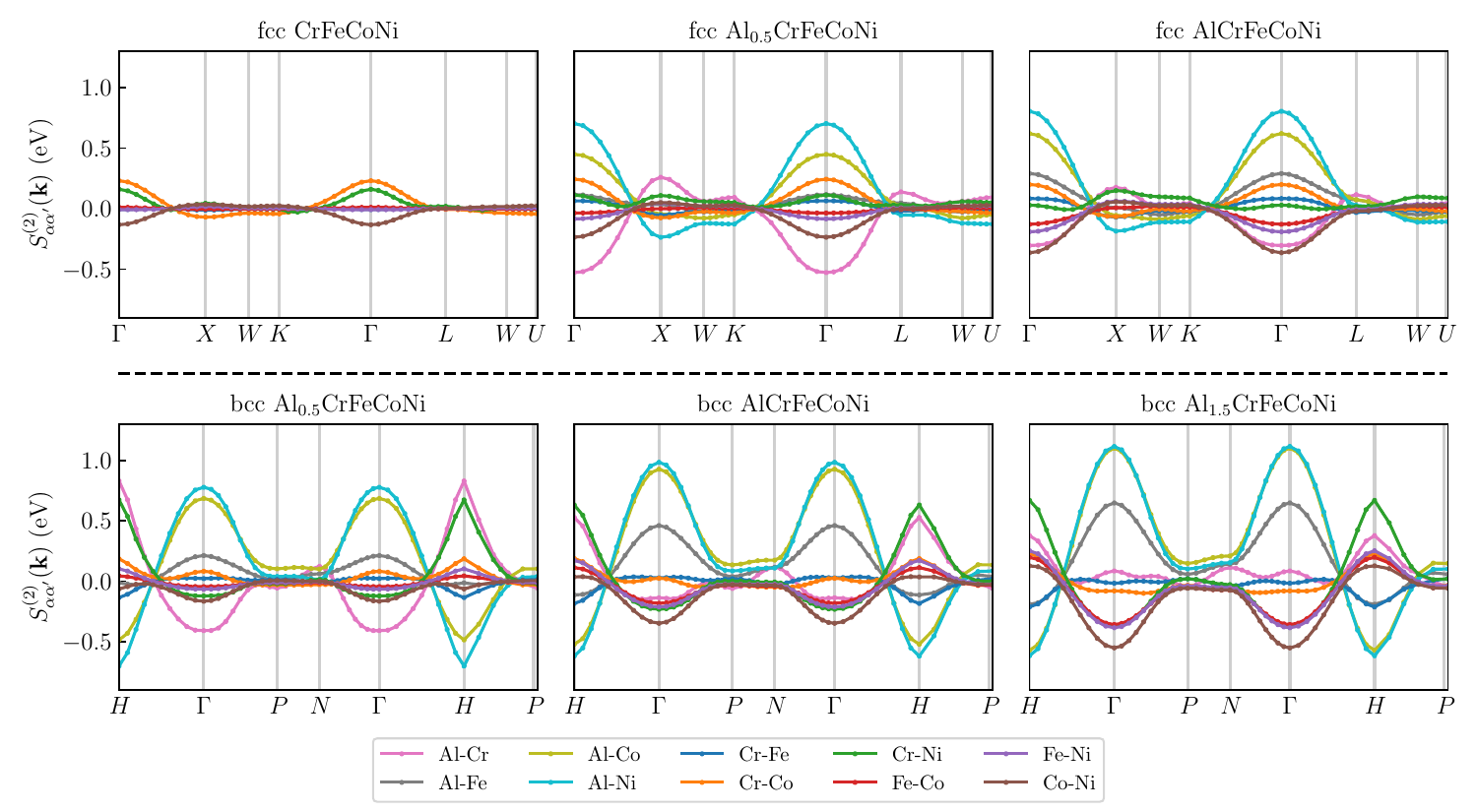}
    \caption{Plots of $S^{(2)}_{\alpha \alpha'}(\mathbf{k})$ for the Al$_x$CrFeCoNi alloy around various high-symmetry lines of the irreducible Brillouin zone for fcc (top row) and bcc (bottom row) lattices. $S^{(2)}_{\alpha \alpha'}(\mathbf{k})$ represents the lattice Fourier transform of an effective pair interaction between chemical species. The profile and relative sizes of of $S^{(2)}_{\alpha \alpha'}(\mathbf{k})$ indicate the likely strength and nature of pairwise atom-atom correlations in the solid solution. It can be seen that the introduction of Al to the system dramatically increases the strength of atom-atom correlations. Interactions are generally stronger when the underlying lattice is bcc compared to fcc.}
    \label{fig:s2_comparison}
\end{figure*}

Proceeding, we consider the electronic DoS for both fcc and bcc structures, as shown in Fig.~\ref{fig:dos_comparison}. The most obvious contrast between the two lattice types is the difference in the contribution to the total DoS made by the narrow, $3d$ bands associated with the transition metals Cr, Fe, Co, and Ni. In the bcc structure, particularly at lower values of $x$, there is a notable dip in the total and species-resolved DoS curves at intermediate energies. This is associated with splitting of $e_g$ (higher energy) and $t_{2g}$ (lower energy) states in this structure. In our calculations, this splitting is increased in the bcc structure compared to the fcc structure. Sample $(l,m)$-resolved DoS plots evidencing this are given in the Supplemental Material~\cite{supplemental}. As for the BSF plots of Fig.~\ref{fig:bsf_comparison}, another key feature is the broad, $sp$ band associated with Al, which can be seen hybridising with the $d$ bands of the transition metals, indicated by localised peaks in its species-resolved DoS. With increasing $x$, a peak in the total DoS at lower energies is maintained in the bcc structure, while in the fcc structure the total DoS appears gradually pushed to higher energies. 

Another important point to note is the decreasing valence electron concentration (VEC) with increasing Al content. For $x=0$  (\textit{i.e.} the composition CrFeCoNi) the VEC is 8.25~$e$/atom, while for $x=1$ (\textit{i.e.} the composition AlCrFeCoNi), the VEC is 7.2~$e$/atom. { Varying VEC has previously been noted as an important factor in determining the underlying lattice type of high-entropy alloys in both experimental~\cite{guo_effect_2011, chen_composition_2018} and theoretical and computational~\cite{poletti_electronic_2014, yang_revisit_2020} studies. For example, Guo \textit{et al.}~\cite{guo_effect_2011} found that, experimentally, alloys with a VEC greater than 8~$e$/atom were typically fcc, while those with a VEC less than 6.87~$e$/atom were typically bcc, findings which are broadly in line with ours in the present study. Attempts have also been made to provide theoretical insight into these experimental observations. Where states at the Fermi surface are free-electron-like in character, for example in alloys of noble metals, the VEC determines where the Fermi sphere intersects with the Brillouin zone boundary~\cite{raynor_progress_1949, poletti_electronic_2014}. As the first Brillouin zones for fcc and bcc lattices are different in shape, it can be argued that the differing shapes of the two Brillouin zones allow for more electrons to be accommodated in lower energy levels for one structure over the other, thus determining the relative stability~\cite{raynor_progress_1949}. However, for systems such as those of the present study, where the $d$-bands are partially filled, such simple arguments do not apply. A more sophisticated argument was made by Poletti and Battezzati~\cite{poletti_electronic_2014}, who employed a modified Friedel model to account for partial filling of $d$-bands. These authors found that, when the VEC was less than 7.5~$e$/atom, the bcc structure was favoured in their modelling, while when the VEC was greater than 7.5~$e$/atom, it was the fcc structure which was favoured. This is in reasonable agreement with our own findings. However, we stress that for systems such as the Al$_x$CrFeCoNi alloy considered in this work, where magnetism plays an important role and the system's $d$-bands are heavily smeared by the disorder, it is challenging to identify a single particular electronic mechanism driving the transition from fcc to bcc lattice type with increasing Al concentration.}

We interpret the combination of the aforementioned band mechanisms (splitting of $3d$ states, $p$-$d$ hybridisation) and reducing VEC as the electronic origins of the stabilisation of the bcc phase with increasing Al content.

In addition to the above features, a comment should also be made about charge transfer between chemical species in the alloyed system. In an alloy, it is typically a good approximation for the average lattice parameter of the system to be determined from a weighted average of the atomic volumes of each element present, in accordance with Vegard's law~\cite{denton_vegards_1991}. However, this means that charge from the Wigner-Seitz cells of `large' atoms in the alloy typically spills over into the Wigner-Seitz cell of the `small' atoms in the alloy. In these systems, we find that Al, the atom with the largest relative atomic volume, typically `loses' some electrons to atoms such as Co and Ni, which have smaller associated atomic volumes. For example, for bcc AlCrFeCoNi ($x=1$), a site occupied by an Al atom (atomic number 13) has an associated charge surrounding it of 12.84 $e$, while Ni (atomic number 28) has an associated charge of 28.13 $e$. (Full details of all relative charge transfers between elements can be found in the potential files in the repository associated with this study.) { Such charge transfer from Al to the other elements present in the composition is reminiscent of the notion of aluminium having a negative effective valency when alloyed with transition metals, as first proposed by Raynor~\cite{raynor_progress_1949}. Such a notion is supported by the propensity of aluminium to form intermetallic compounds with a range of $3d$ transition metals~\cite{stone_negative_1984}.} In previous work, we have found these relative atomic size differences and associated charge transfers between elements to play a key role in determining atomic ordering tendencies as charge rearranges itself in response to an applied chemical perturbation~\cite{woodgate_short-range_2023}. It is therefore vital that these effects are fully considered in the perturbative stability analysis~\cite{khan_statistical_2016, woodgate_modelling_2024}.

\begin{table*}[t]
\begin{ruledtabular}
\begin{tabular}{crccrrrrrc}
Lattice & Composition      & $T_\textrm{ord}$ (K) & $\mathbf{k}_\textrm{ord}$ ($2\pi/a$) & $\Delta$ Al & $\Delta$ Cr & $\Delta$ Fe & $\Delta$ Co & $\Delta$ Ni & Ordered Structure \\ \hline
fcc     & CrFeCoNi         & 380                  & (0,0,1)                              &             & $0.722$     & $-0.049$    & $-0.689$    & $0.016$     & L1$_2$            \\
        & Al$_{0.5}$CrFeCoNi & 562                  & (0,$\frac{1}{2}$,1)                            & $0.803$     & $0.145$     & $-0.167$    & $-0.401$    & $-0.381$    & D0$_{22}$  \\
        & AlCrFeCoNi       & 744                  & (0,$\frac{1}{2}$,1)                            & $0.871$     & $-0.042$    & $-0.195$    & $-0.321$    & $-0.314$    & D0$_{22}$  \\
        & Al$_{1.5}$CrFeCoNi & 987                  & (0,$\frac{1}{2}$,1)                            & $0.882$     & $-0.091$    & $-0.204$    & $-0.301$    & $-0.286$    & D0$_{22}$  \\
        & Al$_2$CrFeCoNi   & 1082                 & (0,0,1)                              & $0.871$     & $-0.040$    & $-0.194$    & $-0.308$    & $-0.328$    & L1$_2$            \\ \hline
bcc     & CrFeCoNi         & 488                  & (0,0,0)                              &             & $0.684$     & $-0.257$    & $0.220$     & $-0.647$    & Phase Seg.        \\
        & Al$_{0.5}$CrFeCoNi & 1982                 & (0,0,1)                              & $0.777$     & $0.173$     & $-0.117$    & $-0.470$    & $-0.363$    & B2                \\
        & AlCrFeCoNi       & 3006                 & (0,0,1)                              & $0.831$     & $0.073$     & $-0.164$    & $-0.404$    & $-0.336$    & B2                \\
        & Al$_{1.5}$CrFeCoNi & 3781                 & (0,0,1)                              & $0.852$     & $0.021$     & $-0.184$    & $-0.374$    & $-0.315$    & B2                \\
        & Al$_2$CrFeCoNi   & 4386                 & (0,0,1)                              & $0.863$     & $-0.012$    & $-0.195$    & $-0.357$    & $-0.298$    & B2               
\end{tabular}
\end{ruledtabular}
\caption{Predicted chemical orderings for the Al$_x$CrFeCoNi system as a function of $x$, as inferred from the described Landau-type linear-response theory for cases with both bcc and fcc as the underlying crystal lattice. Within the concentration wave formalism, a chemical ordering is described by the temperature at which ordering occurs ($T_\text{ord}$), the wave-vector describing ordering ($\mathbf{k}_\text{ord}$), and its \textit{chemical polarisation} ($\Delta c_\alpha$). On the fcc lattice, we find a tendency towards L1$_2$ ordering competing with more complex interactions, which we interpret as a precursor to the experimentally observed D8$_\textrm{b}$ phase. On the bcc lattice, with increasing Al concentration, we find exceptionally strong B2 ordering tendencies, indicating this atomically ordered phase will form directly from the melt, consistent with experimental data.}\label{table:inferred_orderings}
\end{table*}

\begin{figure}[t]
    \centering
    \includegraphics[width=\linewidth]{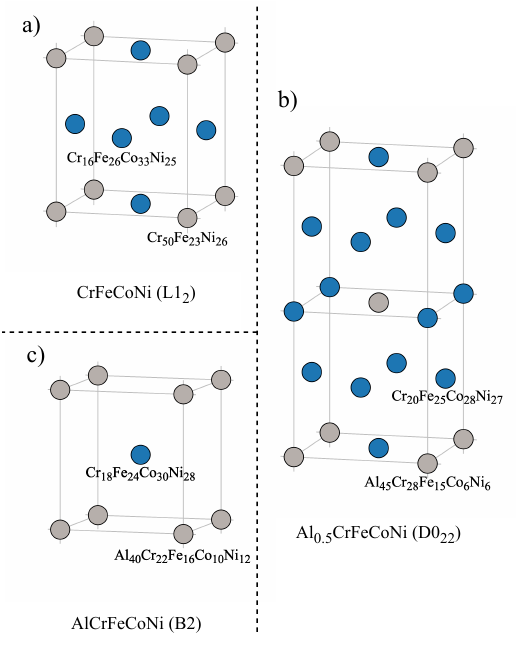}
    \caption{The partially ordered structures for Al$_x$CrFeCoNi predicted by the linear response theory for (a) $x=0$ - an L1$_2$ ordering, (b) $x=0.5$ - a D0$_{22}$ ordering, and (c) $x=1$ - a B2 ordering. We note that in the case of (b), the L1$_2$ and D0$_{22}$ orderings are very close in energy; D0$_{22}$ ordering is \textit{only just} favoured over L1$_2$ in our calculations. { We emphasise that these orderings are only the \emph{initial} orderings inferred from the solid solution by our linear response theory, and that further chemical orderings and/or eventual segregation can occur at lower temperatures. This aspect is examined in Sec.~\ref{sec:atomistic}.}}
    \label{fig:example_structures}
\end{figure}

\subsection{Perturbative Stability Analysis}

We use the self-consistent KKR-CPA one-electron potentials and associated electron densities as the starting point for a linear response calculation assessing the energetic cost of chemical perturbations to the homogeneous, disordered solid solution. { This approach is discussed in detail in Sec.~\ref{sec:chemical_ordering} and in references~~\cite{woodgate_compositional_2022, woodgate_short-range_2023, woodgate_interplay_2023, woodgate_competition_2024, woodgate_modelling_2024, woodgate_integrated_nodate}, so for brevity we only outline the key details here. Chemical} perturbations are naturally expressed within the concentration wave formalism~\cite{khachaturyan_ordering_1978, gyorffy_concentration_1983}. In this approach, we write an inhomogeneous set of site-wise concentrations (Eq.~\ref{eq:partial_occupancies}) as a perturbation to the homogeneous site-wise concentrations (Eq.~\ref{eq:homogeneous_partial_occupancies}),
\begin{equation}
    c_{i\alpha} = c_\alpha + \Delta c_{i\alpha}.
    \label{eq:perturbation}
\end{equation}
In particular, owing to the translational symmetry of the underlying crystal lattice, we write these concentrations in reciprocal space,
\begin{equation}
c_{i\alpha} = c_\alpha + \sum_{\mathbf{k}} e^{i \mathbf{k} \cdot \mathbf{R}_i} \Delta c_\alpha(\mathbf{k}),
\end{equation}
where $\Delta c_\alpha (\mathbf{k})$ represents a static concentration wave with wavevector $\mathbf{k}$ and chemical polarisation $\Delta c_\alpha$. Examples of how these concentration waves can be used to describe various ordered structures imposed on the fcc and bcc lattices are given in Sec.~\ref{sec:theory}. { The wavevectors associated with a given chemical ordering are, naturally, those associated with the superstructure peaks which would be observed in a scattering experiment.}

For a given set of inhomogeneous site-wise concentrations, it is possible to compute the total DFT energy of the system, $E_\textrm{int}(\{c_{i\alpha}\})$,  within the inhomogeneous CPA~\cite{ebert_calculating_2011, faulkner_multiple_2018}. However, for arbitrary ordering wave-vectors and chemical fluctuations, this is computationally expensive, requiring evaluation of a large number of partially ordered supercells. A more elegant and efficient approach is to consider the change of the internal energy of the system in response to an (infinitesimal) perturbation applied to the homogeneous solid solution~\cite{khan_statistical_2016, woodgate_compositional_2022, woodgate_short-range_2023}. It can be shown that first derivatives of this internal energy vanish due to the symmetries of the homogeneous, disordered solid solution, and that the important term is the second derivative,
\begin{equation}
    -\frac{\partial^2 E_\textrm{int}}{\partial c_{i\alpha} \partial c_{j\alpha'}} \equiv S^{(2)}_{i\alpha;j\alpha'},
    \label{eq:s2}
\end{equation}
where we define $S^{(2)}_{i\alpha;j\alpha'}$ for notational convenience. Again, owing to the translational symmetry of the underlying lattice, it is convenient to inspect the lattice Fourier transform of this quantity, writing $S^{(2)}_{\alpha\alpha'}(\mathbf{k})$. {(Relevant definitions of lattice Fourier transforms are provided in Sec.~\ref{sec:fourier_transforms}.)} $S^{(2)}_{\alpha\alpha'}(\mathbf{k})$ is then an $s \times s$ matrix which is a function of the wavevector, $\mathbf{k}$, where $s$ is the total number of chemical species in the alloy. Its evaluation involves self-consistently solving a set of coupled equations in terms of various CPA-related quantities, the most computationally demanding part of which is a convolution integral over the irreducible Brillouin zone (IBZ)~\cite{khan_statistical_2016}. Evaluation of this term has been discussed extensively in earlier works, so we omit these details here for brevity~\cite{khan_statistical_2016, woodgate_compositional_2022, woodgate_short-range_2023, woodgate_interplay_2023, woodgate_competition_2024, woodgate_modelling_2024, woodgate_integrated_nodate}.

Crucially, the $S^{(2)}$s can be directly related to the two-point correlation function, an atomic short-range order (ASRO) parameter~\cite{khan_statistical_2016}. They also can be viewed as the Fourier transforms of pairwise, real-space atom-atom interactions of what is commonly referred to as the Bragg-Williams model~\cite{bragg_effect_1934, bragg_effect_1935}, which has Hamiltonian
\begin{equation}
    H(\{\xi_{i\alpha}\}) = \frac{1}{2}\sum_{i \alpha; j\alpha'} V_{i\alpha; j\alpha'} \xi_{i \alpha} \xi_{j \alpha'},
    \label{eq:b-w}
\end{equation}
where $V_{i\alpha; j\alpha'}$ represents the energy associated with an atom of species $\alpha$ on site $i$ interacting with an atom of species $\alpha'$ on site $j$. It can be shown that, within our formalism, the Fourier transforms of $S^{(2)}_{\alpha\alpha'}(\mathbf{k})$ represent a best choice of effective pair interaction, $V_{i\alpha; j\alpha'}$, for use in this model~\cite{khan_statistical_2016}. {(Note that, on account of the homogeneity of the disordered solid solution, we assume that the real-space interactions are translationally invariant.)} So, by applying an inverse Fourier transform to the $S^{(2)}_{\alpha\alpha'}(\mathbf{k})$ data, we can recover an effective pair interaction for use in subsequent atomistic simulations.

Visualised in Fig.~\ref{fig:s2_comparison} are plots of $S^{(2)}_{\alpha\alpha'}(\mathbf{k})$ around the irreducible Brillouin zone (IBZ) for a range of values of $x$ on both fcc and bcc underlying lattices. It can be seen that, on both lattices, Al-Ni and Al-Co interactions dominate, with other strong interactions between Al-Fe and Co-Ni pairs. We also note that Al-Cr interactions, which are strong at low concentrations of Al, are weakened as Al-content increases, which we understand as originating from changes to the electronic structure of the solid solution from increasing Al-concentration, shown in Fig.~\ref{fig:dos_comparison}. Interactions for the system modelled on the bcc lattice are, on the whole, stronger than the fcc counterpart, suggesting ordering when the system is on the bcc lattice will occur at higher temperatures. We also stress that, for the system modelled without Al present, considered previously in Refs.~\cite{woodgate_compositional_2022, woodgate_interplay_2023}, the relative strength of interactions is substantially reduced, consistent with experimental observations that these systems form stable, single-phase solid solutions down to low temperatures~\cite{yeh_nanostructured_2004, cantor_microstructural_2004}.

\begin{figure*}[p]
    \centering
    \includegraphics[width=\textwidth]{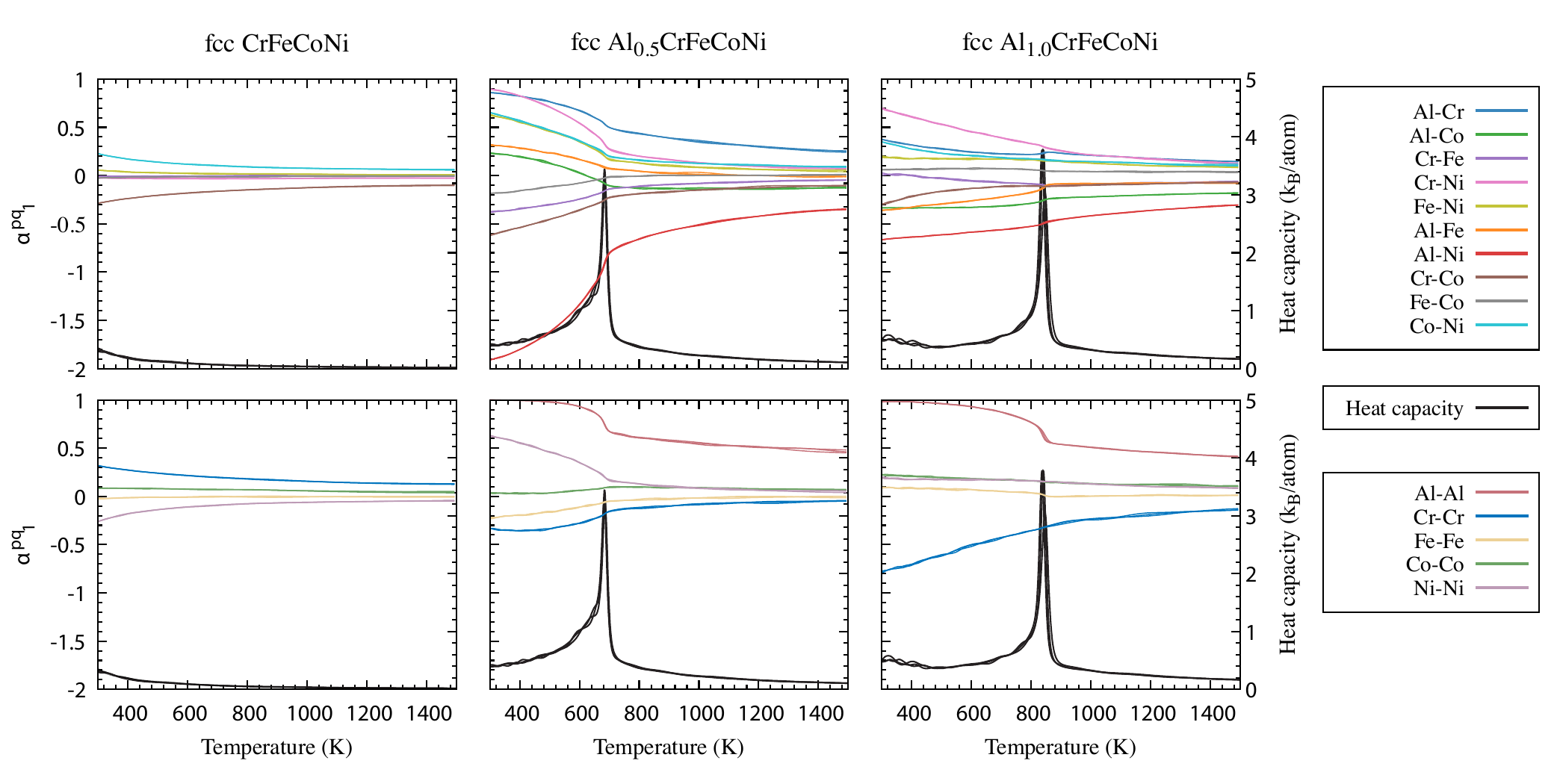}
    \caption{Plots of the isochoric heat capacity (black lines) and Warren-Cowley order parameters (coloured lines) at nearest-neighbour distance for the Al$_x$CrFeCoNi alloy with $x=0, 0.5, 1$ (when the underlying lattice is fcc). Top panels show order parameters for pairs of different atom types, while the bottom panels show parameters for like pairs.}
    \label{fig:fcc_warren-cowleys}
\end{figure*}

\begin{figure*}[p]
    \centering
    \includegraphics[width=\textwidth]{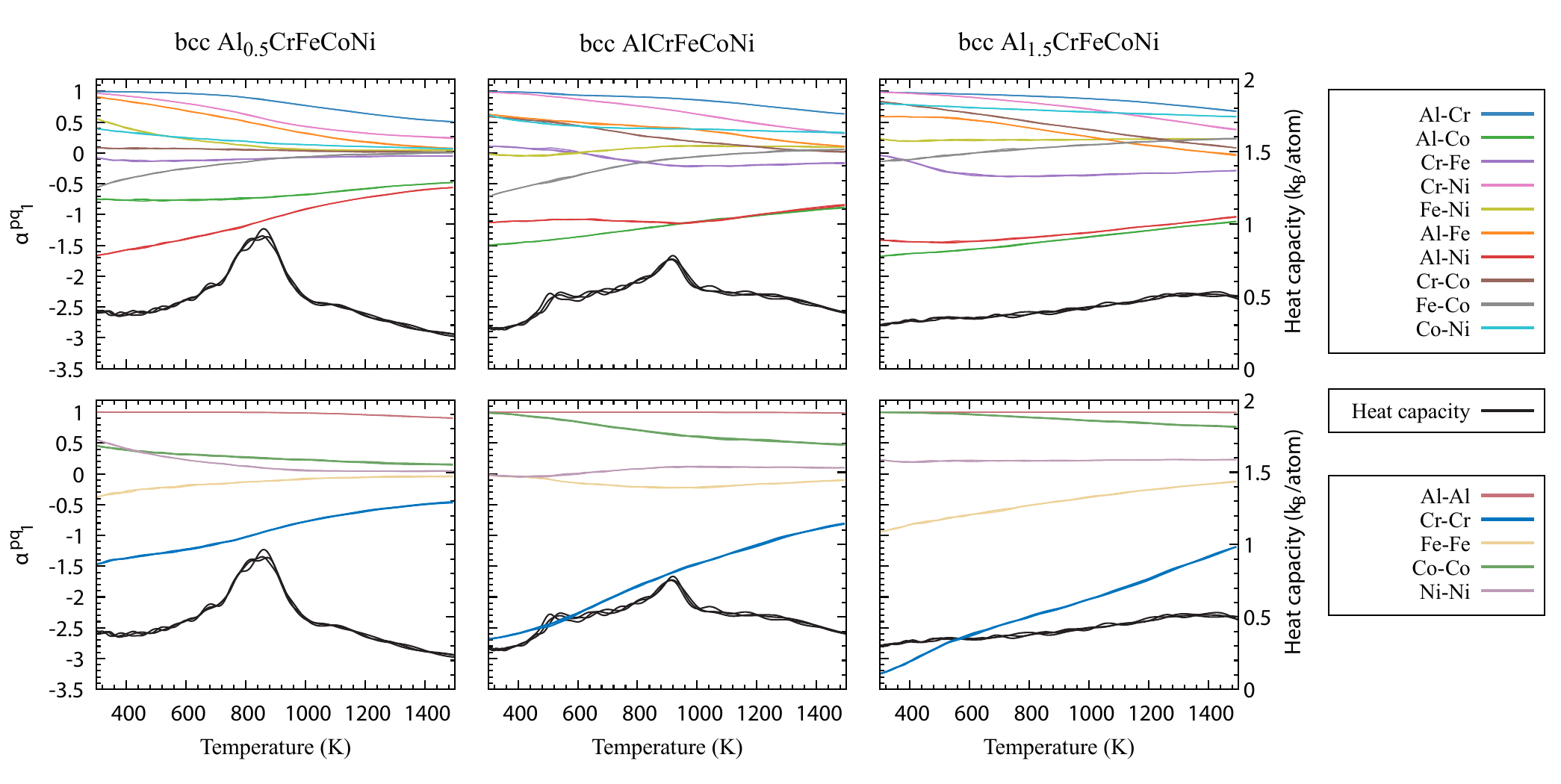}
    \caption{Plots of the isochoric heat capacity (black lines) and Warren-Cowley order parameters (coloured lines) at nearest-neighbour distance for the Al$_x$CrFeCoNi alloy with $x=0.5, 1, 1.5$ (when the underlying lattice is bcc). Top panels show order parameters for pairs of different atom types, while the bottom panels show parameters for alike-pairs. In alignment with Table~\ref{table:inferred_orderings}, at temperatures above those considered in these plots, the system has already exhibited a B2 ordering. For this reason, the Warren-Cowley parameters do not start from zero, rather from finite positive/negative values.}
    \label{fig:bcc_warren-cowleys}
\end{figure*}

\subsection{Inferred Chemical Orderings}
\label{sec:inferred_orderings}

From the reciprocal space $S^{(2)}_{\alpha \alpha'}(\mathbf{k})$ data, it is possible to reconstruct the chemical stability matrix, $\Psi^{-1}_{\alpha\alpha'}(\mathbf{k})$, which assesses the stability of the system to an applied chemical perturbation~\cite{khan_statistical_2016}. This quantity can be thought of as analogous to the Hessian matrix evaluated at a stationary point of the free energy surface of the system. At high temperatures, this matrix is positive definite, and the homogeneous solid solution is stable to any applied chemical perturbation. However, with decreasing temperature, we expect that, at some temperature $T_\text{ord}$ and for some wavevector $\mathbf{k}_\text{ord}$, an eigenvalue of this matrix will pass through zero, and a chemical ordering can be inferred. The wavevector defining the ordering, $\mathbf{k}_\text{ord}$, suggests the ordered structure which will emerge, while the eigenvector itself, $\Delta c_\alpha$, informs us of which chemical species partition themselves onto which sublattice(s). { This aspect is discussed in detail in Sec.~\ref{sec:chemical_ordering}, as well as in earlier works~\cite{woodgate_compositional_2022, woodgate_short-range_2023}.}

Tabulated in \ref{table:inferred_orderings} are inferred ordering temperatures, wavevectors, and concentration wave polarisations for the Al$_x$CrFeCoNi system for a range of values of $x$ for both a bcc and fcc underlying lattice. Some examples of the associated predicted ordered structures are given in Fig.~\ref{fig:example_structures}. We emphasise that, as these ordering temperatures are computed within a single-site theory, they are expected to marginally overestimate the exact transition temperatures~\cite{khan_statistical_2016}. { Further, we stress that these orderings are only the \emph{initial} orderings inferred from the solid solution by our linear response theory, and that further chemical orderings and/or eventual segregation can occur at lower temperatures. This aspect is examined in Sec.~\ref{sec:atomistic}, where atomistic simulations are performed down to room temperature.}

On the fcc lattice, we find that there is close competition between the ordering wavevectors (0, $\frac{1}{2}$, 1) and (0, 0, 1). Alone, the wavevectors (0, 0, 1) and equivalent describe an L1$_2$ ordering~\cite{woodgate_compositional_2022}, visualised in Fig.~\ref{fig:example_structures}. However, the combination of wavevectors (0, 0, 1) and (0, $\frac{1}{2}$, 1) is associated with the D0$_{22}$ structure~\cite{staunton_compositional_1994, khachaturyan_ordering_1978}, also visualised in Fig.~\ref{fig:example_structures}. The D0$_{22}$ structure is closely related to the L1$_2$ structure and is known to emerge in Al-containing alloys~\cite{xu_phase_1990, hong_crystal_1990, lv_deformation_2016}. Formation of the D0$_{22}$ structure requires interactions which are marginally longer-ranged than the L1$_2$ structure, and we suggest that this competition is indicative of more complex ordering tendencies in the system. This is perhaps consistent with the experimental observation of complex D8$_\textrm{b}$ precipitates emerging in this system at low temperatures{, as reported in Ref.~\cite{bloomfield_phase_2022}, which studied phase equilibria by homogenising samples at temperatures of 1200, 1000, 850, and 700~$^\circ$C for a range of values of Al-concentrations, $x$. (We note that} studying emergence of the D8$_\textrm{b}$ phase is beyond the scope of the lattice-based modelling of the present study.) The close competition between L1$_2$ and D0$_{22}$ ordered structures in our results is evidenced by the eigenvalues of the chemical stability matrix as visualised in the Supplemental Material~\cite{supplemental}. Our predicted ordering temperatures for $x \geq 0.5$ are in fair agreement with experimental observations that an L1$_2$ ordered phase emerges below approximately 1000~K~\cite{bloomfield_phase_2022}. The predicted orderings are all driven by Al, indicative of one sublattice rich in Al, with the other sublattice(s) deficient in Al but rich in the other elements. Visualisations of inferred (partially) ordered structures on the fcc lattice are provided in Fig.~\ref{fig:example_structures}.

When the underlying lattice is bcc, for $x \geq 0.5$, our predicted atomic ordering temperatures are all above the melting temperature of the system, which is understood to be approximately 1678~K (= 1405~$^\circ$C) for $x=0.3$ and 1613~K (= 1340~$^\circ$C) for $x=0.7$~\cite{zhang_understanding_2016}. That our predicted ordering temperatures are higher than the melting temperature of the system is consistent with the experimental observation that, for intermediate Al concentrations, no atomically disordered A2 phase emerges and, instead, the system forms the atomically ordered B2 phase straight from the melt~\cite{bloomfield_phase_2022}. Notably, the eigenvectors describing ordering suggest that this B2 ordering is dominated by Al, Co, and Ni, while Fe and Cr atoms remain comparatively disordered and can sit on either sublattice. An example partially ordered structure for the case $x=1$ is visualised in Fig.~\ref{fig:example_structures}. That Al, Co, and Ni atoms order strongly, while Cr and Fe remain comparatively disordered, is in agreement with Ref.~\cite{santodonato_predictive_2018}, which found similar results using a pairwise lattice-based interaction fitted from DFT supercell calculations.

\subsection{Atomistic Simulations}

In a system with many constituent elements and potentially competing chemical interactions, the above partial orderings inferred by the Landau-type theory may not convey a full picture of the phase behaviour of the system. { Indeed, many single-phase high-entropy alloys are expected to eventually become metastable with decreasing temperature.} To address this aspect, we inverse Fourier transform the reciprocal space $S^{(2)}_{\alpha \alpha'}(\mathbf{k})$ data to recover the pairwise, real-space interaction of Eq.~\ref{eq:b-w}. (Our effective pair interactions, $V_{i\alpha; j\alpha'}$, are provided in the repository associated with this work, and are also tabulated in the Supplementary Material.) We find that an interaction limited to the first six coordination shells of the lattice fits the reciprocal space data with extremely good accuracy. Interactions are largest on the first coordination shell and tail off quickly with increasing distance; there are no particularly long-ranged interactions present in our analysis of these systems, although interactions between Al and other elements present are significant up to and including the third coordination shell of the lattice in both fcc and bcc cases. Our fitted interactions are provided in the database associated with this study, and are both tabulated and visualised in the Supplemental Material~\cite{supplemental}.

Using the nested sampling (NS) algorithm, which employs a family of `walkers' to fully explore the potential energy landscape of a given system, we compute the specific heat capacity and atomic short-range order parameters as a function of temperature. The ASRO parameters used in this study are the Warren-Cowley order parameters, $\alpha^{pq}_n$, defined in Sec.~\ref{sec:atomistic}. The symbols $p$ and $q$ are species labels, while $n$ denotes the coordination shell of a particular lattice. When $\alpha^{pq}_n>0$, $p$-$q$ pairs are disfavoured on shell $n$, while when $\alpha^{pq}_n<0$ they are favoured. The value 0 corresponds to the ideal, random, solid solution.

The fcc lattice heat capacities and nearest neighbour order parameters calculated by NS---shown in Fig.~\ref{fig:fcc_warren-cowleys}---show that there is no ordering transition above room temperature when Al is excluded, reflecting the role that Al plays in driving order. When Al is added to the system, first order phase transitions can be seen at approximately 680 and 920 K  for $x=0.5$ and $1$ respectively, in reasonable agreement with the inferred ordering temperatures recorded in Table~\ref{table:inferred_orderings}. (Note that the predicted ordering temperatures of Table~\ref{table:inferred_orderings} are obtained within a mean-field theory, so are naturally expected to slightly overestimate atomic ordering temperatures.)

\begin{figure}[t]
    \centering
    \includegraphics[width=0.45\textwidth]{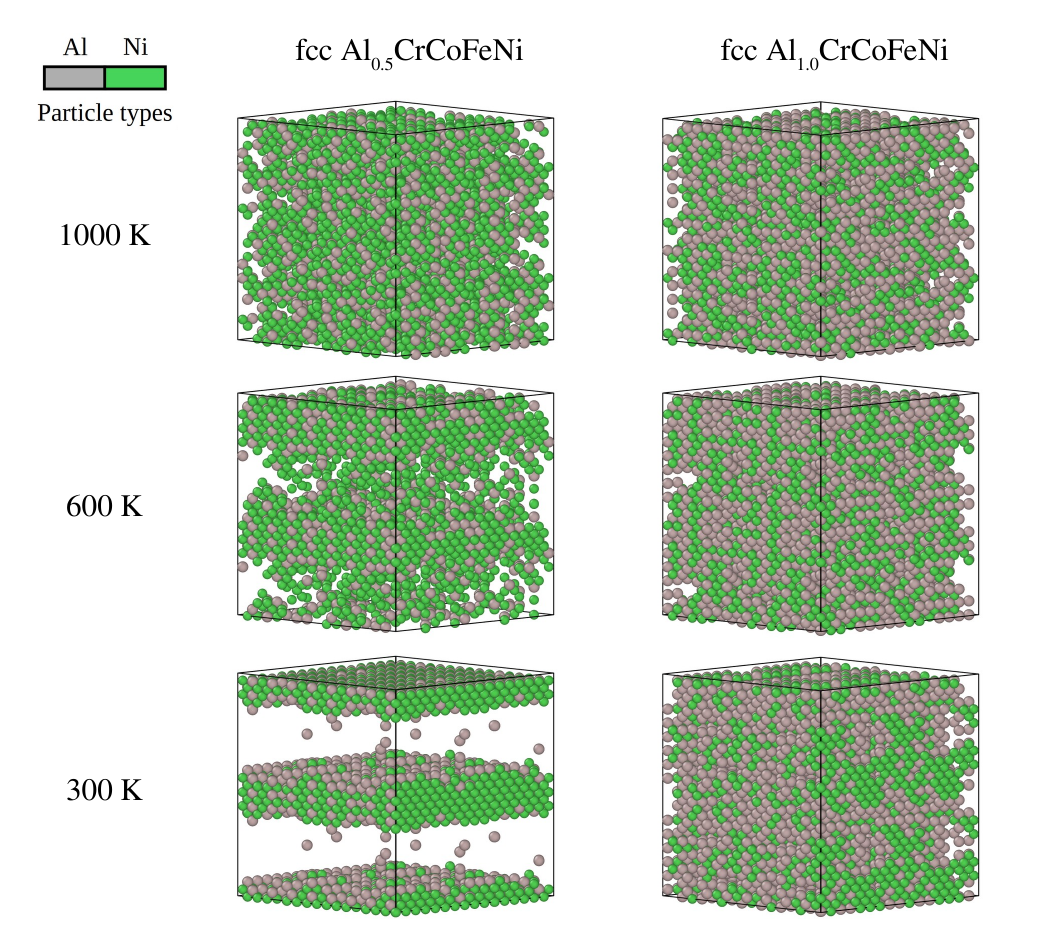}
    \caption{Atomic configurations generated by NS for Al$_x$CoCrFeNi on the fcc lattice ($x=0.5$ and $1$), at temperatures of $300, 600$ and $1000$~K. We only show the Al (grey) and Ni (green) atoms in these snapshots in order to emphasise the phase separation between AlNi and the other constituent elements. It can be seen that, when $x=0.5$, Al and Ni are able to form stoichiometric D0$_{22}$, encouraging them to separate from Fe, Co and Cr. In contrast, in the $x=1$ alloy, Al prefers to evenly disperse throughout the lattice along with Ni. Note that these configurations have been replicated using periodic boundary conditions $2\times2\times2$ times, for clarity. Images generated using Ovito~\cite{stukowski_visualization_2010}.}
    \label{fig:fcc_snapshots}
\end{figure}

In the case of $x=0.5$  the transition corresponds with an increase in Al-Ni and Ni-Ni correlation, as L1$_2$- and D0$_\text{22}$-type ordering begins to manifest. This transition also coincides with the onset of phase separation between Al-Ni and the other alloy constituents, as shown by the decrease in correlation between all Al-$X$ and Ni-$X$ pairs ($X=$Co, Cr, Fe). This phase separation can be seen in the left hand panels of Fig.~\ref{fig:fcc_snapshots}, where stoichiometric Ni$_3$Al-D0$_{22}$ can be seen to form in layers as temperature decreases. Given that the system has an overall Ni$_2$Al composition, there is an excess of Al at the phase boundaries.

\begin{figure}[t]
    \centering
    \includegraphics[width=0.45\textwidth]{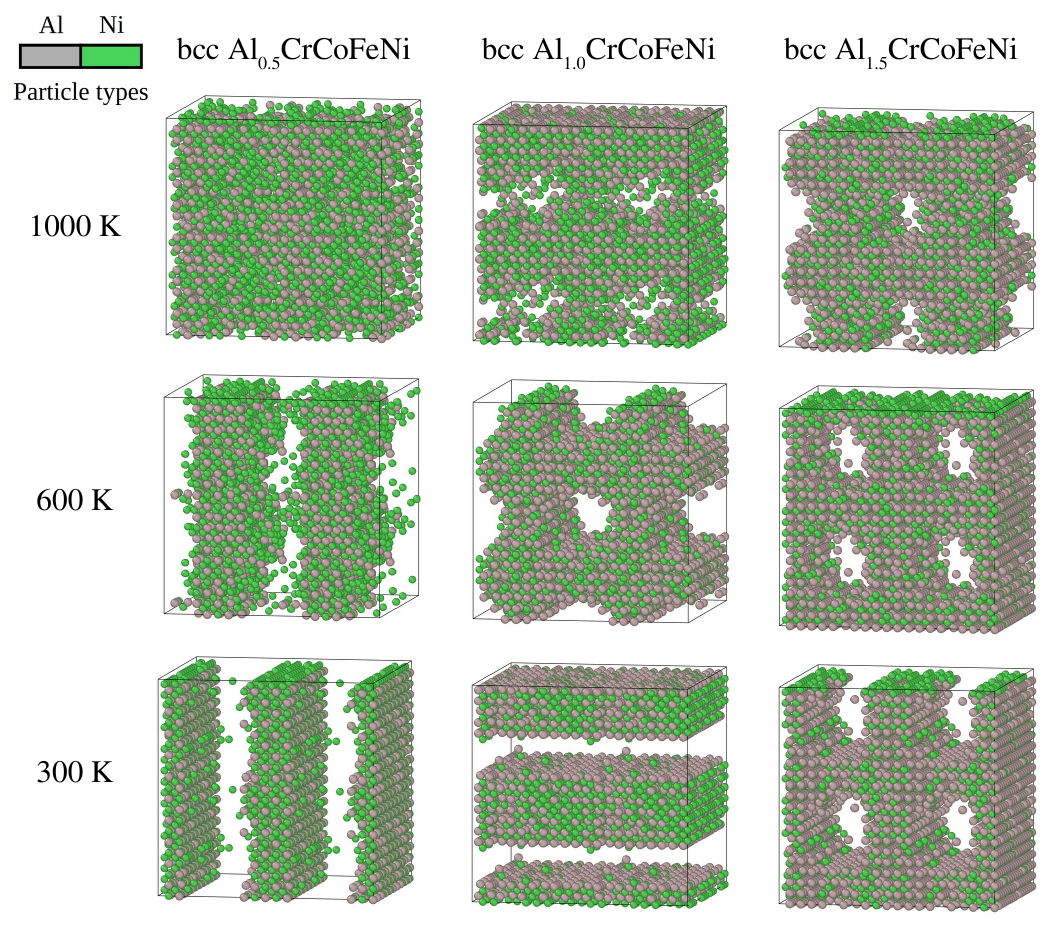}
    \caption{Atomic configurations generated by NS for Al$_x$CoCrFeNi on the bcc lattice ($x=0.5, 1$ and $1.5$), at temperatures of $300, 600$ and $1000$~K. We only show the Al (grey) and Ni (green) atoms in these snapshots in order to emphasise the phase separation between B2 AlNi and the other constituent elements. They show the different forms of phase separation that occur depending on Al content: planar and cylindrical. Note that these configurations have been replicated using periodic boundary conditions $2\times2\times2$ times, for clarity. Images generated using Ovito~\cite{stukowski_visualization_2010}.}
    \label{fig:bcc_snapshots}
\end{figure}

When $x=1$, Al and Ni no longer form the D0$_{22}$ phase separately from the other constituents. Instead, Al orders on a D0$_{22}$ lattice that is dispersed throughout the system, in accordance with the inferred chemical orderings depicted in panel b of Fig.~\ref{fig:example_structures}. As a result, Ni also diffuses across the system as it is energetically encouraged to bond with Al, as the right-hand panels of Fig.~\ref{fig:fcc_snapshots} show. Hence we find less dramatic changes in nearest neighbour order parameters over the transition compared to the $x=0.5$ system.

Figure~\ref{fig:bcc_warren-cowleys} depicts heat capacity and ASRO parameter data for the bcc lattice with $x=0.5, 1$ and $1.5$. As predicted by the inferred B2 transition temperatures ($T_\text{ord}\geq 1982$~K for $x\geq0.5$), the Al-Al order parameters have the maximal value of $1$ below $1500$~K, as the Al atoms have already been distributed onto next-nearest neighbour sites. 

The peaks in heat capacity correspond to the process of phase separation between B2 AlNi and the rest of the system, as shown by  the NS-generated atomic configurations displayed in Fig.~\ref{fig:bcc_snapshots}, where Al and Ni atoms have been isolated for clarity. We find that different types of separation dominate depending on Al content and temperature. When $x=0.5$, phases separate in a planar fashion, and when $x=1.5$ they separate such that the CoCrFe matrix is contained by the AlNi atoms in a pseudo-cylindrical fashion. In the case of $x=1$, these two modes are energetically comparable at $600$ K such that NS generates configurations with both modes. Below $500$ K, the temperature at which the heat capacity suddenly decreases, we only observe the planar mode, indicating that the change in separation-type is a second order phase transition.

The above results, as well as the results inferred directly via the concentration wave analysis in Sec.~\ref{sec:inferred_orderings}, are broadly consistent with other theoretical studies, in addition to experimental works. Ref.~\cite{farkas_model_2020} used embedded atom model (EAM) potentials and Monte Carlo simulations to equilibrate structures at a temperature of 500~K. These authors found short-range order followed by an L1$_2$ ordering driven by Al with increasing Al-concentration when the underlying lattice was fcc, and a strong B2 ordering driven by Al when the underlying lattice was bcc. Ref.~\cite{hasan_short-range_2024} found similar SRO tendencies using another EAM potential, finding Al-Fe and Cr-Cr nearest-neighbour pairs favoured, and Al-Al and Al-Cr pairs disfavoured. In Ref.~\cite{santodonato_predictive_2018}, lattice-based Monte Carlo simulations using nearest-neighbour pair interactions were employed, and it was found that, when the underlying lattice was bcc, their simulations decomposed into B2 Al-Ni-Co rich phases and disordered bcc Cr-Fe phases. Similarly, Ref.~\cite{anber_role_2022} reports combined experimental and theoretical results, {where experimental samples for a range of Al concentrations are annealed at temperatures between 1250 and 700~$^\circ$C. These authors find} that the system decomposes into B2 Ni-Al and Cr-rich phases when the underlying lattice is bcc { for temperatures between 900 and 700~$^\circ$C}. That our results are in alignment with these earlier studies reassures us of the predictive capabilities of our modelling approach.

\section{Discussion}
\label{sec:conclusions}

In summary, results have been presented analysing the phase behaviour of the Al$_x$CrFeCoNi high-entropy alloy within a DFT-based modelling framework. We have considered the underlying crystal lattice, the nature of atomic short-range order, as well as emergent atomic long-range order. In good agreement with experiment, we find that for $x>1$, the bcc lattice is favoured over fcc, while for $x<0.5$, it is fcc that is preferred. In the region $0.5\leq x \leq 1$, the two lattice types are very close in energy to one another, in alignment with experimental observations of phase coexistence~\cite{kao_microstructure_2009, bloomfield_phase_2022}.

Via a perturbative analysis of the internal energy of the disordered alloy as evaluated within the KKR-CPA, we have examined the dominant atom-atom correlations in the solid solution. For both lattice types, we find the strongest correlations are between Al and the other elements present, although, with increasing Al-concentration, Ni-Co pairs also interact strongly. Application of a Landau-type theory to a perturbative expansion of the free energy of the system enables chemical orderings to be inferred directly. When the underlying lattice is fcc, we predict either an L1$_2$ or D0$_{22}$ ordering dominated by Al. In the L1$_2$ structure,  Al preferentially sits on the corners of a simple cubic lattice, while the other elements remain disordered on the other sublattices. Similarly, the D0$_{22}$ ordering is also driven by Al preferentially moving to one sublattice. When the underlying lattice is bcc, we predict a B2 ordering at temperatures higher than the melting temperature of the alloy, consistent with experimental observations that this ordered phase forms directly from the melt. This B2 ordering is driven by Al, with one sublattice predicted to be rich in Al, and the other rich in Ni and Co. Fe and Cr are predicted to be spread more homogeneously.

The perturbative analysis also enables extraction of a simple, pairwise model of the internal energy of the alloy, suitable for study via lattice-based atomistic simulations. Here the thermodynamics of each alloy are investigated using the nested sampling algorithm, which efficiently samples different atomic arrangements while estimating their individual contributions to the global partition function. This approach provides greater accuracy and detail in describing thermodynamic phase behaviour compared to mean-field methodologies. We find results in good agreement with the earlier linear-response analysis, along with emergent multi-phase behaviour where different forms of AlNi-FeCoCr phase separation dominate or coexist as temperature changes. The advantage of using nested sampling over Markov chain Monte Carlo is demonstrated by its ability to sample multiple thermodynamically competitive phases simultaneously. 

These results serve both to demonstrate the validity of this computationally efficient modelling approach, as well as to give fundamental physical insight into the experimentally observed phase behaviour of this technologically relevant HEA. We are able both to infer the nature of potential, emergent, ordered structures, and to predict which chemical species partitions onto which sublattice. In addition, as the methodology is based on DFT calculations of the electronic structure of the disordered solid solutions, we can elucidate the underlying electronic mechanisms driving ordering tendencies. These results are therefore of relevance to both experimentalists and theorists, and facilitate improved understanding of the phase behaviour of the Al$_x$CrFeCoNi system.

\section{Methods}
\label{sec:theory}

\subsection{The $\mathbf{S^{(2)}}$ theory for multicomponent alloys}
\label{sec:chemical_ordering}

The $S^{(2)}$ theory for multicomponent alloys as used in this work represents a perturbative analysis of the response to the CPA reference medium to an applied inhomogeneous chemical perturbation~\cite{khan_statistical_2016, woodgate_compositional_2022}.  The approach uses a Landau-type expansion of the free energy of the system, as evaluated within the CPA, about a homogeneous (disordered) reference state to obtain the two-point correlation function, an ASRO parameter, {\it ab initio}. This perturbative analysis can be thought of as analogous to density-functional perturbation theory used for obtaining lattice dynamics, {\it e.g.} phonon dispersion relations, within DFT calculations~\cite{baroni_phonons_2001}. We stress that effects of the response of the electronic structure and the rearrangement of charge due to the applied chemical perturbation are fully included. The methodology is the natural extension of an earlier modelling approach developed for binary alloys~\cite{staunton_compositional_1994, johnson_first-principles_1994, clark_van_1995}, the groundings of which lie in statistical physics and in seminal papers on concentration waves authored by Khachaturyan~\cite{khachaturyan_ordering_1978} and Gyorffy and Stocks~\cite{gyorffy_concentration_1983}. Full details of the theory, its implementation, and extensive discussion can be found in a range of earlier works~\cite{khan_statistical_2016, woodgate_compositional_2022, woodgate_short-range_2023, woodgate_interplay_2023, woodgate_competition_2024, woodgate_modelling_2024, woodgate_integrated_nodate}.

\begin{figure}[t]

    \centering
    \includegraphics[width=\linewidth]{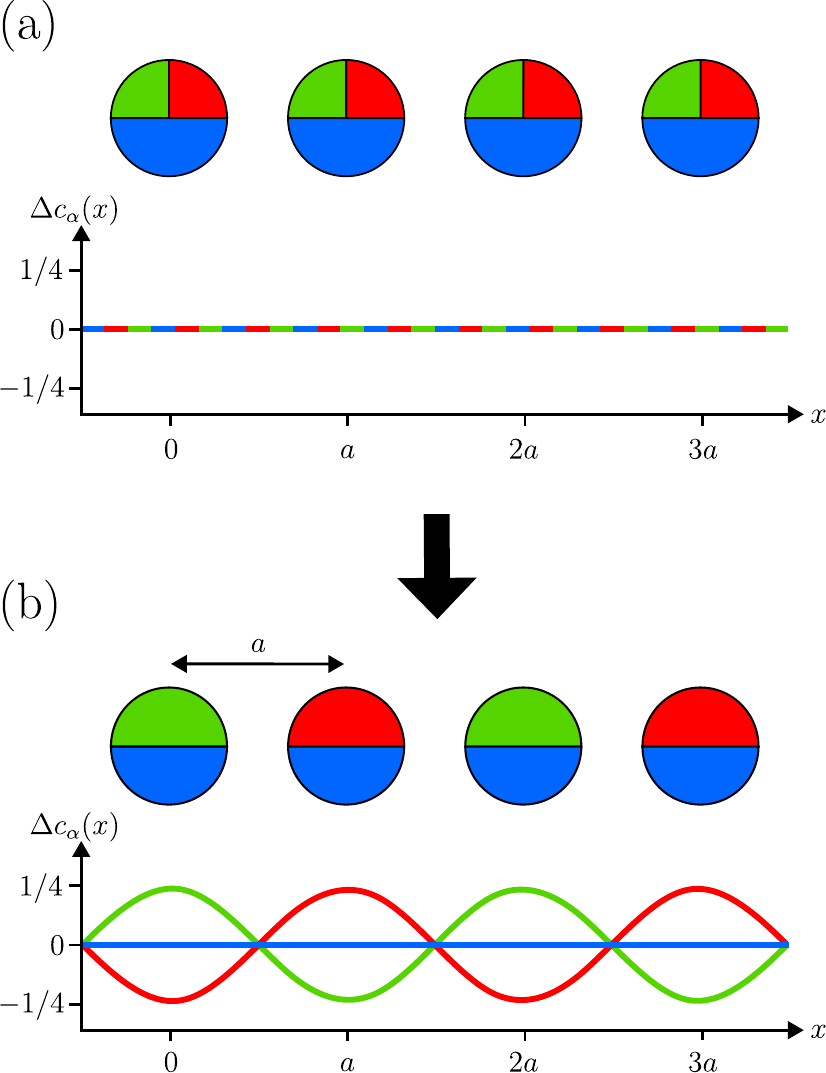}
    \caption{Illustration of a concentration wave modulating partial atomic site occupancies in a toy, 1D, 3-component alloy. The top panel, (a), shows the homogeneous alloy with no perturbation applied, while the bottom panel, (b), shows the alloy with an example applied chemical perturbation. The alloy's overall (average) concentrations are $c_\alpha = (1/2, 1/4, 1/4)$, with species coloured blue, green, and red respectively. Real-space lattice sites are at positions $R_i = ia$ The normalised chemical polarisation of the example applied concentration wave is $\Delta c_\alpha/\| \Delta c_\alpha \| = (0, 1/\sqrt{2}, -1\sqrt{2})$, indicating that partial occupancies of blue atoms remain unchanged, while partial occupancies of green and red atoms are changed by equal and opposite amounts. The associated wave vector is $k=\pi/a$.}
    \label{fig:concentration_wave_illustration}
\end{figure}

The methodology assumes a fixed, ideal lattice, either face-centred cubic (fcc) or body-centered cubic (bcc) for the alloys studied in this paper. This lattice represents the averaged atomic positions in the solid solution---a valid approximation in high-entropy alloys where local lattice distortions are known to be small. A complete description of the atomic configuration of a substitutional alloy with this fixed underlying lattice is given by the site occupation numbers, $\{\xi_{i\alpha}\}$, where 
\begin{equation}
\xi_{i\alpha} :=
\begin{cases}
1 , \; \text{if site $i$ is occupied by an atom of species $\alpha$}\\
0 ,\; \text{otherwise}.
\end{cases}
\end{equation}
To ensure that each lattice site has one (and only one) atom occupying it, we impose the constraint that $\sum_\alpha \xi_{i\alpha}$=1 for all lattice sites $i$. The total concentration of a chemical species, $c_\alpha$ is given by $c_\alpha = \frac{1}{N} \sum_i \xi_{i\alpha}$, where $N$ is the total number of lattice sites in the system. The most natural choice of atomic long-range order parameter is given by the ensemble average of the site occupancies, $c_{i\alpha} = \langle \xi_{i\alpha} \rangle$, where the $\{c_{i\alpha}\}$ are referred to as the site-wise concentrations. In the limit of high temperature, where the solid solution is completely disordered, these occupancies will be spatially homogeneous and take the values of the overall concentration of each chemical species, {\it i.e.} $\lim_{T \to \infty} c_{i\alpha} = c_\alpha$. However, below any atomic disorder-order transition, the site occupancies are expected to acquire a spatial dependence. The inhomogeneous set of site-occupancies can be written as a fluctuation to the concentration distribution of the homogeneous system, $c_{i\alpha} = c_\alpha + \Delta c_{i\alpha}$. Given the translational symmetry of the underlying crystal lattice, it is natural to write these fluctuations in reciprocal space using a concentration wave formalism, as suggested by Khachaturayan~\cite{khachaturyan_ordering_1978}. We therefore write
\begin{equation}
c_{i\alpha} = c_\alpha + \sum_{\mathbf{k}} e^{i \mathbf{k} \cdot \mathbf{R}_i} \Delta c_\alpha(\mathbf{k})
\end{equation}
to describe a chemical fluctuation, where $\mathbf{R}_i$ is the lattice vector with corresponding occupancy $c_{i\alpha}$. 

\begin{figure}[b]
    \centering
    \includegraphics[width=\linewidth]{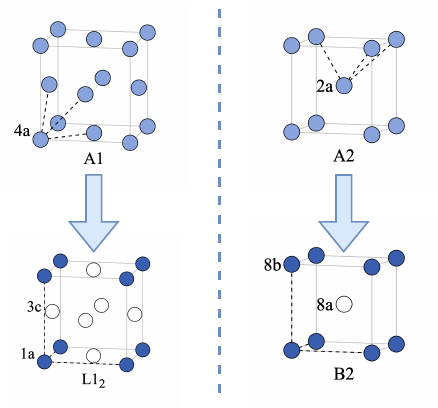}
    \caption{Examples of ordered structures which can be described by concentration waves. The L1$_2$ structure is an ordered structure on the fcc lattice, while the B2 structure is an ordered structure on the bcc lattice. Both are described by wavevectors $\mathbf{k} = (0,0,2\pi/a)$ and equivalent for their respective lattice types. Structures are described by their \emph{Strukturbericht} designation, primitive lattice vectors are denoted by dashed lines, and atomic sites are given their Wyckoff labels.}
    \label{fig:concentration_wave_examples}
\end{figure}

{An illustration of how this scheme can describe a partially-ordered superstructure is given in Fig.~\ref{fig:concentration_wave_illustration} for a toy, 1D, 3-component alloy. Then, as an example of how this formalism can be used to describe ordered structures imposed on a three-dimensional crystal lattice,} in Fig.~\ref{fig:concentration_wave_examples} we show the L$1_2$ order imposed on the fcc lattice for an, $A_3B$ binary system, $c_\alpha = (\frac{3}{4}, \frac{1}{4})$. The L$1_2$ ordered structure, represented by atoms of species $B$ at the corners of the unit cell and atoms of species $A$ at the face-centres, is described by $\mathbf{k}_\text{ord} = (0,0,1)$ and equivalent, the (normalised) change in concentration $\Delta c_\alpha = \frac{1}{\sqrt{2}} (-1, 1)$. {Of course, the wavevector(s) describing the ordering, $\mathbf{k}_\text{ord}$ are naturally related to the superstructure peaks that would be observed in a scattering experiment~\cite{khachaturyan_ordering_1978}.}

This formalism is naturally applicable to multicomponent alloys. As an example, consider a B2 ordering in a Heusler compound with composition $A_2BC$. (The B2 structure is visualised in Fig.~\ref{fig:concentration_wave_examples}.) An ordering where atoms of species $A$ occupy the 8a sites, while atoms of species $B$ and $C$ remain disordered on the 8b sites, would be described by a wavevector $\mathbf{k} = (0,0,1)$ and equivalent (as for a binary B2 ordering) but with a (normalised) change in concentration $\Delta c_\alpha = \frac{1}{\sqrt{6}} (2, -1, -1)$.

Above any chemical disorder-order transition temperature the natural choice of atomic short-range order parameter is then the two point correlation function,
\begin{equation}
\Psi_{i\alpha j \alpha'} = \langle \xi_{i\alpha}\xi_{j\alpha'} \rangle - \langle \xi_{i\alpha} \rangle \langle \xi_{j\alpha'}\rangle,
\end{equation}
which quantifies the strength and range of atom-atom correlations between chemical species. The two-point correlation function is, of course, related to the energetic cost of chemical fluctuations~\cite{khan_statistical_2016}; the dominant ASRO will be associated by those chemical fluctuations which are least energetically costly.

To assess the energetic cost of a given chemical fluctuation, we approximate the free energy, $\Omega$, of an alloy with an inhomogeneous set of partial lattice site occupancies, $\{c_{i\alpha}\}$, by
\begin{equation}
\begin{aligned}
    \Omega^{(1)}[\{c_{i\alpha}\}] =& -\frac{1}{\beta} \sum_{i\alpha} c_{i\alpha} \ln c_{i\alpha} \\
    &- \sum_{i\alpha} \nu_{i\alpha} c_{i\alpha} + \langle \Omega_\text{el} \rangle_0 [\{c_{i\alpha}\}],
    \label{eq:free_energy2}
\end{aligned}
\end{equation}
where the three terms on the right-hand side of Eq.~\ref{eq:free_energy2} describe entropic contributions, from site-wise chemical potentials, $\{ \nu_{i\alpha}\}$, and an average of the electronic contribution to the free energy of the system, respectively. We then expand this free energy about the homogeneous reference state ({\it i.e.} the disordered solid solution) in terms of the applied inhomogeneous fluctuation $\{\Delta c_{i\alpha}\}$. This is a Landau-type series expansion and is written
\begin{align}
    \Omega^{(1)}[\{c_{i\alpha}\}] &= \Omega^{(1)}[\{c_{\alpha}\}] + \sum_{i\alpha} \frac{\partial \Omega^{(1)}}{\partial c_{i\alpha}} \Big\vert_{\{c_{\alpha}\}} \Delta c_{i\alpha} \nonumber \\ 
    &+ \frac{1}{2} \sum_{i\alpha; j\alpha'} \frac{\partial^2 \Omega^{(1)}}{\partial c_{i\alpha} \partial c_{j\alpha'}} \Big\vert_{\{c_{\alpha}\}} \Delta c_{i\alpha}\Delta c_{j\alpha'} + \dots.
\label{eq:landau}   
\end{align}
In the full linear response theory, the site-wise chemical potentials of Eq.~\ref{eq:free_energy2} serve as Lagrange multipliers to conserve the overall concentrations of each chemical species. However, the variation of these multipliers is understood to be irrelevant to the underlying physics, so terms involving these derivatives are dropped~\cite{khan_statistical_2016, woodgate_compositional_2022, woodgate_short-range_2023}. In addition, combined with the translational symmetry of the disordered solid solution, the requirement that overall concentrations of each chemical species be conserved ensures that the first-order term of Eq.~\ref{eq:landau} is zero. Keeping terms to second-order, the change in energy due to an applied chemical perturbation is written
\begin{equation}
    \delta \Omega^{(1)} = \frac{1}{2} \sum_{i\alpha; j\alpha'} \Delta c_{i\alpha} [\beta^{-1} \, C_{\alpha\alpha'}^{-1} - S^{(2)}_{i\alpha, j\alpha'}] \Delta c_{j\alpha'}.
\label{eq:chemical_stability_real}
\end{equation}
The first term in square brackets, $C_{\alpha \alpha'}^{-1} = \frac{\delta_{\alpha \alpha'}}{c_\alpha}$, is a diagonal, positive definite matrix is associated with entropic contributions to the free energy, while the second term, $-\frac{\partial^2 \langle \Omega_\text{el} \rangle_0}{\partial c_{i\alpha} \partial c_{j\alpha'}} \equiv S^{(2)}_{i\alpha;j\alpha'}$ is the second-order concentration derivative of the average energy of the disordered alloy as evaluated within the CPA. Evaluation of this term amounts to self-consistently solving a ring of coupled equations in terms of various CPA-relevant quantities, carefully incorporating the rearrangement of the electrons due to the applied chemical perturbation. This set of coupled equations are defined in Ref.~\cite{khan_statistical_2016}, and their solutions first examined and discussed in Ref.~\cite{woodgate_compositional_2022}.

Within the outlined concentration wave formalism, $S^{(2)}_{i\alpha;j\alpha'}$ is evaluated in reciprocal space, and therefore the change in free energy of Eq.~\ref{eq:chemical_stability_real} is written accordingly as:
\begin{equation}
    \delta \Omega^{(1)} = \frac{1}{2} \sum_{\bf k} \sum_{\alpha, \alpha'} \Delta c_\alpha({\bf k}) [\beta^{-1} C^{-1}_{\alpha \alpha'} -S^{(2)}_{\alpha \alpha'}({\bf k})] \Delta c_{\alpha'}({\bf k}).
\label{eq:chemical_stability_reciprocal}
\end{equation}
The matrix in square brackets $[\beta^{-1} C^{-1}_{\alpha \alpha'} -S^{(2)}_{\alpha \alpha'}({\bf k})]$, referred to as the chemical stability matrix, is related to an estimate of the ASRO, $\Psi_{i\alpha;j\alpha'}$. It can be thought of as analogous to a Hessian matrix of second derivatives evaluated at a stationary point of the free energy landscape. When searching for an disorder-order transition, we start with the high temperature solid solution, where all eigenvalues of this matrix are positive and the system is stable to applied chemical perturbations. We then progressively lower the temperature and look for the point at which the lowest lying eigenvalue of this matrix passes through zero for any $\mathbf{k}$-vector in the irreducible Brillouin Zone. When this eigenvalue passes through zero at some temperature $T_\text{us}$ and wavevector $\mathbf{k}_\text{us}$, the disordered solid solution is unstable to that applied perturbation and we infer the presence of an disorder-order transition with chemical polarisation $\Delta c_\alpha$ given by the associated eigenvector. In this fashion we can predict both dominant ASRO and also the temperature at which the solid solution becomes unstable and a chemically ordered phase emerges.

\subsection{Pairwise Atomistic Model}
\label{sec:atomistic}

The linear response theory provides information about the initial chemical ordering in the system, but it is possible to go further and map derivatives of the internal energy of the disordered alloy, $S^{(2)}_{\alpha\alpha'}(\mathbf{k})$, to a pairwise real-space interaction. When an appropriate sampling technique is applied to this model, the phase behaviour can be studied in detail and equilibrated, lattice-based configurations extracted for illustration and further study. As an example, we highlight the work of Ref.~\cite{shenoy_collinear-spin_2024}, which used lattice-based configurations with physically motivated atomic arrangements in the training dataset of a machine-learned interatomic potential to study a variety of physical properties of the prototypical austenitic stainless steel---Fe$_7$Cr$_2$Ni.

The real-space model discussed above, referred to as the Bragg-Williams model~\cite{bragg_effect_1934, bragg_effect_1935}, is lattice-based and has a Hamiltonian of the form
\begin{equation}
    H = \frac{1}{2}\sum_{i \alpha; j\alpha'} V_{i\alpha; j\alpha'} \xi_{i \alpha} \xi_{j \alpha'}.
    \label{eq:b-w2}
\end{equation}
The effective pairwise interactions, $V_{i\alpha; j\alpha'}$, are recovered from $S^{(2)}_{\alpha\alpha'}(\mathbf{k})$ by means of a inverse Fourier transform. The mapping from reciprocal-space to real-space and fixing of the gauge degree of freedom on the $V_{i\alpha; j\alpha'}$ is specified in earlier works~\cite{khan_statistical_2016, woodgate_compositional_2022, woodgate_short-range_2023}. These interactions are assumed to be isotropic and the sum in Eq.~\ref{eq:b-w} is then taken as a sum over coordination shells, {\it i.e.} first-nearest neighbours, second-nearest neighbours, \textit{etc.} 

\subsection{Nested Sampling Algorithm}
\label{sec:nested_sampling}

The nested sampling (NS) method was originally introduced in the area of Bayesian statistics~\cite{bib:skilling,bib:skilling2,NS_all_review} and has since been adapted to efficiently sample the different ways in which atomistic systems may be arranged~\cite{1st_NS_paper, NS_mat_review}, allowing for automatic generation of atomic configurations that are thermodynamically relevant at different temperatures. 
Thanks to its iterative top-down approach, NS does not require any prior knowledge of a system's important structures, or the location of phase transitions within configuration space, thus making the algorithm truly unbiased and fully predictive. 
The efficacy of NS has been thoroughly demonstrated via the calculation of various materials' pressure-temperature phase diagrams~\cite{pt_phase_dias_ns,NS_lithium,ns_zirconium}. It has often uncovered systems' thermodynamically stable phases that were previously unknown~\cite{jagla}, and has recently improved the accuracy of a machine learned interatomic potential by identifying important structures for inclusion in its training set~\cite{ns_carbon}. 

NS uses a set of configurations, usually referred to as \emph{walkers} or the \emph{live set}, to iteratively sample the atomic many-body potential energy surface. The initial set of $K$ walkers is always generated randomly to represent the high-energy gas phase configurations. 
In the current work the potential energy is defined only on a fixed lattice, and thus our configuration space is restricted to different relative arrangements of the various atom types on the predefined lattice sites. Therefore, we generated the initial configurations by placing the elements in the required proportions but randomly, either on the fcc or bcc lattice sites. 
During the iterative exploration, the highest energy configuration is picked from among the $K$ walkers in each NS cycle. The phase space volume bound by the corresponding energy contour is estimated as $K/(K+1)$-th fraction of the previous contour.
The highest energy configuration is then replaced in the walker set, with the criteria that the new configuration has to be uniformly randomly picked and its energy has to be lower than that of the one being replaced. 
The simplest way of finding a suitable such new configuration is to use rejection sampling. However, as the sampling progresses, the probability of randomly generating a lower energy arrangement of elements quickly diminishes. Thus, we use another strategy: we clone an existing walker and perform a series of Monte Carlo steps (with every step accepted if the energy remains below the limit) until we can assume the configuration to be sufficiently decorrelated. 
In case of most stochastic explorations, usually the step-size is adjusted to maintain a required acceptance ratio during the random walk. 
Since the lattice is fixed in our model, our random walk steps can only include swaps of different atom types, with no parameter to adjust. This means that fewer and fewer attempted swaps are accepted as we sample lower energy regions, with no means of improving the acceptance ratio. To counteract this, we increase the number of attempted swaps from each iteration if the number of accepted steps per walk drops below 200.  

The nested sampling iterations are repeated until the set of walker configurations reach the region of the phase space that is low enough in energy to contribute down to the temperature range we are interested in.

It has to be noted that to be able to estimate the phase space volume of each energy contour, the algorithm requires us to be able to sort the walkers in strictly monotonous order. Hence, no two configurations can have the exact same energy. This is of particular concern since the space we explore is discrete in energy, and it is possible to have two different arrangements of the various elements on the lattice sites resulting in the same total energy.
To avoid this, we add a small random number of up to $10^{-8}$~Ry to the energy of the configuration when the walker is first generated. This does not affect the calculated thermodynamic properties, but allows us to always uniquely pick the highest energy walker, without violating the requirement that the samples are uniformly distributed.  

A unique advantage of NS is that it provides an easy access to the partition function at an arbitrary temperature during the post-processing of the generated sample configurations. 
Using the energies of the discarded walkers and the corresponding estimates for the phase space volume, we can calculate the canonical partition function, $Q$ as a function of number of particles $N$, volume $V$, and temperature $T$, via the following summation,
\begin{equation}
    Q(N,V,T)=\sum_i (\Gamma_{i-1}-\Gamma_{i}) e^{-\beta E_i},    
\label{eq:part_func}
\end{equation}
where $E_i$ is the energy of the configuration discarded in the $i$-th NS iteration, $\Gamma_i=(K/K+1)^i$ is the phase space volume of the configuration space bound by this, and $\beta$ is the thermodynamic temperature.
From the partition function we can easily evaluate thermodynamic response function, such as the heat capacity, to help locate phase transitions,
\begin{equation}
    C_V (T)=\frac {\partial}{\partial T}\Big( -\frac{\partial \ln Q (N,V,T)} {\partial \beta}\Big)_{N,V}.   
\label{eq:cv}
\end{equation}
We can also calculate the phase space weighted averages of observables (\textit{e.g.} the radial distribution function) to evaluate their finite temperature expected values using
\begin{equation}
    \langle A \rangle \approx \frac{1}{Q}\sum_i A_i (\Gamma_{i-1}-\Gamma_{i}) e^{-\beta E_i},    
\label{eq:average}
\end{equation}
where $A_i$ is the value of the observable at the $i$-th NS iteration.

\subsection{Computational Details}
\label{sec:computational_details}

We use the all-electron HUTSEPOT code~\cite{hoffmann_magnetic_2020} to generate self-consistent, one electron potentials within the KKR formulation~\cite{ebert_calculating_2011, faulkner_multiple_2018} of density functional theory (DFT)~\cite{martin_electronic_2004}. Chemical disorder is described within the coherent potential approximation (CPA)~\cite{soven_coherent-potential_1967, faulkner_calculating_1980, johnson_total-energy_1990}, while the the paramagnetic state of the alloys is modelled within the disordered local moment (DLM) picture~\cite{pindor_disordered_1983, staunton_disordered_1984, gyorffy_first-principles_1985}. We perform spin-polarised, scalar-relativistic calculations within the atomic sphere approximation (ASA)~\cite{temmerman_atomic_1978}. We use an angular momentum cutoff of $l_\text{max} = 3$ for basis set expansions, a $20\times20\times20$ $\mathbf{k}$-point mesh for integrals over the Brillouin zone, and a 24 point semi-circular Gauss-Legendre grid in the complex plane to integrate over valence energies. We use the local spin-density approximation (LSDA) and the Perdew-Wang exchange-correlation functional~\cite{perdew_accurate_1992}. The linear response results are obtained from an computational implementation of the theory described in Section~\ref{sec:chemical_ordering} and  discussed in detail in earlier works~\cite{khan_statistical_2016, woodgate_compositional_2022, woodgate_short-range_2023, woodgate_interplay_2023, woodgate_competition_2024, woodgate_modelling_2024, woodgate_integrated_nodate}.

Lattice parameters are set to be consistent with their experimental values~\cite{wang_effects_2012}, and are tabulated in the Supplemental Material~\cite{supplemental}. For values of $x$ where the lattice type is not observed experimentally ({\it e.g.} bcc, $x=0$; fcc, $x=2$) we perform a volume-conserving transformation from fcc to bcc structures (or vice-versa) to obtain a lattice parameter. For intermediate values of $x$ we interpolate linearly between lattice parameters in accordance with Vegard's law. For $x=0$, the experimental fcc lattice parameter is 3.57~\AA, while for $x=2$, the experimental bcc lattice parameter is 2.88~\AA~\cite{wang_effects_2012}. In a previous study, we have found that the results of the $S^{(2)}$ theory are not particularly sensitive to small deviations in lattice parameters around their experimental values~\cite{woodgate_competition_2024, woodgate_modelling_2024}.

Nested sampling calculations were carried out on fcc and bcc supercells of size $8\times8\times8$ fcc cubic unit cells (2048 atoms) and $10\times10\times10$ bcc cubic unit cells (2000 atoms) respectively.
To assess the method's convergence of thermally averaged properties, we repeatedly carried out three parallel runs on a single system while increasing the number of walkers. This way, we determined that $12,000$ walkers was enough to achieve the desired agreement between independent runs. 
To ensure that the algorithm produced configurations that would contribute to thermally averaged properties down to room temperature, we found that at least $(12.0 \pm 0.5) \times10^6$ iterations were required, taking 30 hours to complete using a single processor. 
To sufficiently de-correlate a new configuration from its parent, each iteration initially consisted of 3000 atom swaps, and by the end of the sampling process this was gradually increased up to 12000 swaps to account for the decreasing acceptance rate. 
The code implementing the nested sampling algorithm used in this work is freely available online via the DOI \href{https://doi.org/10.5281/zenodo.10808347}{10.5281/zenodo.10808347}.

{
\subsection{Lattice Fourier Transforms}
\label{sec:fourier_transforms}

In sections~\ref{sec:chemical_ordering} and \ref{sec:atomistic}, lattice Fourier transforms are invoked to switch between real- and reciprocal-space representations of quantities as needed. We define these transformations explicitly below. In all cases, we consider a system with $N$ Bravais lattice sites. The vector for the real-space position of lattice site $i$ is denoted $\mathbf{R}_i$. A scalar quantity $f_i$ transforms according to
\begin{equation}
    f(\mathbf{k}) = \frac{1}{\sqrt{N}} \sum_i e^{-i\mathbf{k}\cdot\mathbf{R}_i} f_i,
\end{equation}
while the inverse transformation is defined via
\begin{equation}
    f_i = \frac{1}{\sqrt{N}} \sum_i e^{i\mathbf{k}\cdot\mathbf{R}_i} f(\mathbf{k}).
\end{equation}
A pairwise real-space quantity $A_{ij}$, such as the effective pair interaction, $V_{i\alpha;j\alpha'}$ of Eq.~\ref{eq:b-w2}, transforms according to
\begin{equation}
    A(\mathbf{k}) = \frac{1}{\sqrt{N}} \sum_i e^{-i\mathbf{k}\cdot\mathbf{R}_i} A_{i0},
\end{equation}
where we have assumed translational invariance. The inverse transformation is then defined via
\begin{equation}
    A_{ij} = \frac{1}{\sqrt{N}} \sum_i e^{i\mathbf{k}\cdot(\mathbf{R}_i - \mathbf{R}_j}) A(\mathbf{k}).
\end{equation}
Note that the underlying lattice types considered in this work (fcc and bcc) have a monatomic basis, and there is therefore no need to include a sum over basis vectors in any of the above transformations.
}

\section*{Data Availability}
\label{sec:data_availability}
The data produced for this study are available through Zenodo via the DOI \href{https://doi.org/10.5281/zenodo.10972974}{10.5281/zenodo.10972974}. This repository includes self-consistent KKR-CPA potentials, reciprocal-space $S^{(2)}$ data, real-space effective pair interactions, and NS-generated atomic configurations. Specific questions should be directed to the corresponding author(s).

\section*{Code Availability}
\label{sec:code_availability}
The all-electron HUTSEPOT code~\cite{hoffmann_magnetic_2020} used for constructing the self-consistent one-electron potentials of DFT within the KKR-CPA is available at \href{https://hutsepot.jku.at/}{hutsepot.jku.at}. The code implementing the $S^{(2)}$ theory for multicomponent alloys~\cite{khan_statistical_2016} is available from C.D.W. or J.B.S. upon reasonable request. The code for performing lattice-based simulations implementing the nested sampling algorithm to study alloy phase behaviour using the obtained atom-atom interchange parameters is freely available on Zenodo via the DOI \href{https://doi.org/10.5281/zenodo.10808347}{10.5281/zenodo.10808347}.

\begin{acknowledgments}
C.D.W. was supported by a studentship within the UK Engineering and Physical Sciences Research Council-supported Centre for Doctoral Training in Modelling of Heterogeneous Systems, Grant No. EP/S022848/1. J.B.S. and C.D.W. acknowledge support from EPSRC Grant EP/W021331/1. L.B.P. acknowledges support from the EPSRC through the individual Early Career Fellowship, Grant No. EP/T000163/1. Computing facilities were provided by the University of Warwick's Scientific Computing Research Technology Platform (SCRTP). \\
\end{acknowledgments}

\section*{Author Contributions}
\label{sec:author_contributions}
C.D.W. and J.B.S. conceived of the approach. J.B.S. led development of the  code for the ASRO linear response calculations. C.D.W. and L.B.P. developed the code implementing the nested sampling algorithm applied to the pairwise atomistic Hamiltonian. Self-consistent DFT calculations were performed by C.D.W. The linear response calculations evaluating ASRO {\it ab initio} were performed by J.B.S. and C.D.W., with subsequent analysis and recovery of pairwise atomistic model performed by C.D.W. Atomistic simulations and subsequent analysis were performed by G.A.M. The first draft of the manuscript was written by C.D.W. and subsequently input was received from all authors.

\section*{Competing Interests}
\label{sec:competing_interests}
{ J.B.S. acts as an associate editor for npj Computational Materials. The other authors declare no competing interests.}

\end{document}